\documentclass[journal]{IEEEtran}
\usepackage{cite}
\usepackage{amsmath,amssymb,amsfonts}
\usepackage{algorithmic}
\usepackage{algorithm}
\usepackage{graphicx}
\usepackage{textcomp}
\usepackage{makecell}
\usepackage{subfigure}
\usepackage[caption=false,font=normalsize,labelfont=sf,textfont=sf]{subfig}
\usepackage{booktabs}
\usepackage{empheq}
\usepackage{balance}
\usepackage{tabularx}
\usepackage{multirow}
\usepackage{booktabs}
\usepackage[table,xcdraw]{xcolor}

\begin{document}

\title{A Survey on the Symbiosis of Artificial Intelligence and Terahertz Networks}
\author{\IEEEauthorblockN{Chong Han,~\IEEEmembership{Senior Member,~IEEE}, Jingting Jiang, Meixia Tao,~\IEEEmembership{Fellow,~IEEE}, and Wenjun Zhang,~\IEEEmembership{Fellow,~IEEE}}
\thanks{Chong Han is with the Terahertz Wireless Communications (TWC) Interdisciplinary Research Center, Shanghai Jiao Tong University, Shanghai 200240, China, and also with the Cooperative Medianet Innovation Center (CMIC), Shanghai Jiao Tong University, Shanghai 200240, China (e-mail: chong.han@sjtu.edu.cn).}
\thanks{Jingting Jiang is with the Terahertz Wireless Communications (TWC) Interdisciplinary Research Center, Shanghai Jiao Tong University, Shanghai 200240, China (e-mail: jingting.jiang@sjtu.edu.cn).}
\thanks{Meixia Tao and Wenjun Zhang are with the School of Information Science and Electronic Engineering, Shanghai Jiao Tong University, Shanghai 200240, China (e-mail: mxtao@sjtu.edu.cn; zhangwenjun@sjtu.edu.cn).}

}
\date{\today}
\maketitle

\begin{abstract}
The Terahertz (THz) band (0.1--10 THz) has emerged as a critical frontier for future wireless systems, offering ultra-wide bandwidths for Terabit-per-second (Tbps) communications together with high-resolution sensing. However, practical THz deployment faces distinctive challenges, including complex channel characteristics, high-dimensional and large-scale optimization, hardware constraints, and highly dynamic network environments. Artificial Intelligence (AI) provides powerful capabilities for modeling, signal processing, optimization, prediction, and real-time decision-making, offering a transformative approach to addressing these challenges. Reciprocally, the unprecedented bandwidth and high-resolution sensing capabilities of THz systems provide a promising physical infrastructure for AI, supporting AI training and inference, physical-world data acquisition, and emerging physical computing paradigms.

This survey presents a systematic and comprehensive study of this bidirectional relationship, progressing from \textit{AI for THz} to \textit{THz for AI} and, ultimately, toward an \textit{AI--THz symbiosis}. We first establish the foundations of AI--THz integration by introducing key AI techniques and evaluation platforms, followed by AI-assisted THz hardware design and channel modeling. We then systematically investigate how AI can enhance THz systems across physical-layer functionalities, higher-layer network protocols, and emerging services including edge computing and sensing. Reversing the perspective, we examine how THz can enable next-generation AI systems through ultra-high-capacity connectivity for AI training and inference, high-resolution sensing for multi-modal and Physical AI, and THz-enabled physical computing. Finally, we identify open challenges and future research directions toward deeper AI--THz integration. By bridging intelligent networking with high-performance communication and high-resolution sensing, this survey envisions AI and THz as mutually reinforcing technologies for future wireless networks and emerging AI systems.

\end{abstract}


\section{Introduction}
\label{intro}

\subsection{Background and Applications}
The Terahertz (THz) band, broadly spanning 0.1--10 THz, represents one of the last major frontiers of the electromagnetic (EM) spectrum that remains largely underexploited for information technologies~\cite{akyildiz2022terahertz}. Its significance extends far beyond providing additional spectrum for 6G or future generations of wireless communications~\cite{giordani20206g,tataria20216g}. The enormous available bandwidth can enable Terabit-per-second (Tbps) connectivity, while the extremely short wavelengths and distinctive interactions between THz waves and matter provide unprecedented sensing capabilities. These properties position the THz spectrum as a strategic physical resource at the convergence of \textit{communication, sensing, and intelligence}. As conventional lower-frequency spectrum becomes increasingly congested and physically constrained, exploiting the THz band opens a fundamentally new design space for future information systems~\cite{survey_thz,tera_ref}. In particular, the rapid evolution of artificial intelligence (AI), from data-intensive foundation models to embodied and Physical AI, is creating unprecedented demands for extreme-capacity connectivity and high-fidelity physical-world perception. This evolution further elevates the role of THz from a spectrum resource for wireless communications to a potential physical infrastructure for future AI systems.

\begin{figure*}[!htbp]
    \centering
    \includegraphics[width=0.8\linewidth]{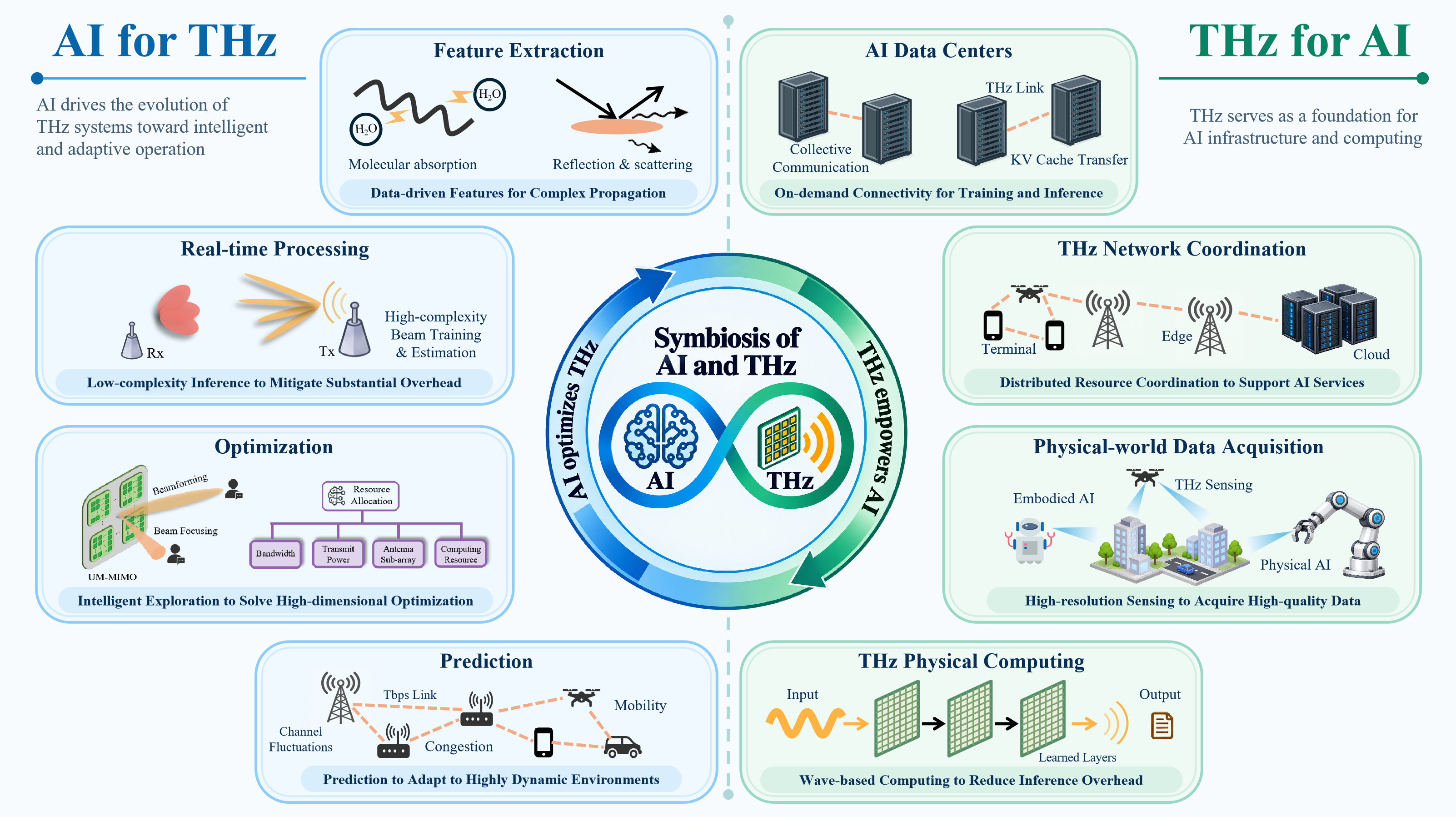}
    \caption{Symbiosis of AI and THz.}
    \label{fig_aithz}
    \vspace{-3mm}
\end{figure*}

THz can support a broad range of emerging applications, with THz and AI increasingly serving as mutual enablers.

\begin{itemize}
    \item \textbf{From terrestrial to aerial and space communications:} 
    In terrestrial networks, THz communications can support high-capacity fronthaul and backhaul links~\cite{IEEE8021532023,kokkoniemi2025dynamic}, as well as ultra-high-speed data delivery in short-range hotspot scenarios such as kiosk downloading~\cite{IEEE8021532023,he2017stochastic}. Beyond terrestrial environments, THz links can further establish high-speed inter-satellite and air-to-ground connections, supporting space-air-ground integrated networks (SAGINs)~\cite{gao2025terahertz,THz_air_ground,THz_space}. As these networks become increasingly heterogeneous and dynamic, AI-driven optimization and network management become essential for managing their complexity~\cite{hu2023deep,hu2024tera,hu2025graph}.

\item \textbf{From communications to sensing and ISAC:} 
Short THz wavelengths also enable sub-millimeter resolution sensing for localization, imaging, and material identification. This distinctive property provides a compelling foundation for integrated sensing and communication (ISAC) \cite{han2024thz}, where the same THz infrastructure can simultaneously connect intelligent agents and enable them to perceive their physical surroundings \cite{xiang2026adaptive}. Such communication--sensing integration is particularly important for embodied and Physical AI systems that must continuously sense, understand, and interact with complex real-world environments.

\item \textbf{From THz communications to THz for AI:} More fundamentally, the rapid evolution of AI is creating communication requirements that may extend well beyond those envisioned for conventional mobile broadband. 
In AI data centers, large model training, Agentic AI inference, and Physical AI simulations generate intensive rack-to-rack traffic, potentially demanding wireless links beyond 100~Gbps. Reconfigurable THz links can support these data exchanges through direct high-performance connections, thereby enabling wireless data center architectures for AI workloads~\cite{han2025wires}.
Additionally, autonomous driving requires low-latency exchange of sensor data and perception information for real-time decision-making~\cite{sakaguchi2021towards}. Emerging generative video communications~\cite{zhang2025generative} require transport of high-fidelity and increasingly AI-generated multi-modal content. 
More broadly, AI-intensive applications such as the Metaverse and extended reality (XR)~\cite{wu2024physical,zugno2025use} further illustrate the convergence of extreme-rate communications, sensing, and computing. 
These requirements are difficult to accommodate through incremental improvements within conventional spectrum alone, motivating the exploitation of the enormous bandwidth available at THz frequencies. 
In this sense, THz is not merely a higher-frequency extension of 6G communications but has the potential to become a key wireless infrastructure for AI.
\end{itemize}

\subsection{Challenges and Motivation}
Despite the remarkable potential of THz communication for future wireless ecosystems, unique physical properties introduce distinct technical challenges. 
First, pronounced molecular absorption \cite{han2022molecular} and diffuse scattering \cite{sheikh2019study} lead to complex frequency-dependent propagation behaviors, making accurate analytical modeling intractable. 
Second, THz system design entails substantial computational overhead. The large dimensionality of ultra-massive MIMO (UM-MIMO) architectures and a more apparent near-field effect \cite{han2024cross} increase the complexity of channel estimation and beamforming. 
Third, from the networking perspective, multi-dimensional resource allocation becomes increasingly difficult as the coordination of massive antenna arrays and fragmented spectrum for numerous concurrent users exceeds the optimization capabilities of conventional algorithms \cite{hu2025graph}.
Fourth, directional THz beams \cite{petrov2020capacity}, which are particularly sensitive to user mobility and blockage, together with rapidly varying traffic conditions \cite{akyildiz2014teranets}, lead to highly dynamic network environments and pose challenges to efficient network management and stable link performance.

To address these challenges, conventional solutions, based on deterministic mathematical models or iterative algorithms, often struggle to capture the nonlinear characteristics of THz systems or efficiently handle high-dimensional problems under rapidly varying network conditions. In contrast, AI provides a promising approach to overcome these limitations and optimize network performance. 
AI techniques, especially deep learning (DL), can learn complex nonlinear mappings from massive datasets to facilitate accurate channel characterization and estimation for the THz band. Furthermore, DL can also provide low-complexity inference for computationally intensive tasks such as hardware EM simulation and beam management. 
For large-scale beamforming and resource allocation problems, learning-based optimization, particularly deep reinforcement learning (DRL), enables real-time decision-making and achieves superior performance. 
Moreover, AI provides unique predictive capacities. The analysis of spatial and temporal correlations enables precise forecasts of channel fluctuations, traffic patterns, and user mobility, granting THz networks essential adaptability in dynamic environments.

\begin{table*}[!htbp]
\centering
\caption{Comparison of AI for Communications Surveys}
\label{tab_ref_survey}
\setlength{\tabcolsep}{3pt}
\renewcommand{\arraystretch}{1.1}
\setlength{\aboverulesep}{0.3ex}
\setlength{\belowrulesep}{0.3ex}

\newcommand{\covfull}{\ensuremath{\checkmark}}
\newcommand{\covpart}{\ensuremath{\circ}}
\newcommand{\covnone}{--}

\begin{tabularx}{\textwidth}{
    @{}
    >{\centering\arraybackslash}p{2.7cm}
    >{\centering\arraybackslash}p{3.4cm}
    *{11}{>{\centering\arraybackslash}X}
    @{}
}
\toprule

\multirow{2}{*}{\makecell[c]{\textbf{Technical}\\\textbf{Domain}}}
& \multirow{2}{*}{\makecell[c]{\textbf{Research}\\\textbf{Topics}}}
& \multicolumn{6}{c}{
    \makecell[c]{\textbf{AI for Wireless Communications}}}
& \multicolumn{4}{c}{
    \makecell[c]{\textbf{AI for THz Communications}}}
& \multirow{3}{*}{
    \makecell[c]{\textbf{This}\\\textbf{Survey}}}
\\
\cmidrule(lr){3-8}
\cmidrule(lr){9-12}

&
& \cite{zhang2021deep}
& \cite{tang2021survey}
& \cite{ye2024artificial}
& \cite{Sharma2025Overview}
& \cite{cui2025overview}
& \cite{zhu2025wireless}
& \cite{boulogeorgos2021machine}
& \cite{khan2023machine}
& \cite{farhad2023terahertz}
& \cite{yu2024ai}
&
\\

\midrule

\makecell[c]{\textbf{Simulators}}
& \makecell[c]{Simulators/platforms}
& \covnone & \covpart & \covnone & \covnone
& \covpart & \covpart & \covnone & \covpart
& \covfull & \covpart & \covfull
\\
\midrule

\multirow{2}{*}{
    \makecell[c]{\textbf{AI for THz}\\\textbf{Hardware Design}}}
& RF/antenna design
& \covnone & \covnone & \covnone & \covnone
& \covnone & \covnone & \covnone & \covnone
& \covpart & \covnone & \covfull
\\
\cline{2-13}
& Impairment mitigation
& \covpart & \covnone & \covpart & \covnone
& \covnone & \covnone & \covfull & \covnone
& \covnone & \covfull & \covfull
\\
\midrule

\multirow{2}{*}{
    \makecell[c]{\textbf{AI for THz}\\\textbf{Channels}}}
& \makecell[c]{Channel modeling}
& \covfull & \covnone & \covnone & \covnone
& \covpart & \covpart & \covnone & \covnone
& \covnone & \covfull & \covfull
\\
\cline{2-13}
& Channel estimation
& \covfull & \covnone & \covfull & \covpart
& \covfull & \covfull & \covfull & \covpart
& \covpart & \covfull & \covfull
\\
\midrule

\multirow{4}{*}{
    \makecell[c]{\textbf{AI for PHY Layer}\\\textbf{Functionality}}}
& Coding/modulation
& \covfull & \covnone & \covfull & \covpart
& \covpart & \covfull & \covfull & \covnone
& \covfull & \covpart & \covfull
\\
\cline{2-13}
& Beamforming
& \covpart & \covnone & \covfull & \covpart
& \covpart & \covpart & \covfull & \covpart
& \covfull & \covfull & \covfull
\\
\cline{2-13}
& Beam management
& \covnone & \covnone & \covpart & \covpart
& \covfull & \covpart & \covfull & \covfull
& \covfull & \covpart & \covfull
\\
\cline{2-13}
& End-to-end transceiver
& \covfull & \covnone & \covfull & \covnone
& \covpart & \covfull & \covpart & \covnone
& \covnone & \covnone & \covfull
\\
\midrule

\multirow{3}{*}{
    \makecell[c]{\textbf{AI for Higher}\\
                 \textbf{Layer Networking}}}
& MAC layer
& \covnone & \covfull & \covpart & \covfull
& \covfull & \covfull & \covfull & \covnone
& \covfull & \covpart & \covfull
\\
\cline{2-13}
& Network layer
& \covnone & \covfull & \covnone & \covpart
& \covpart & \covpart & \covfull & \covnone
& \covfull & \covnone & \covfull
\\
\cline{2-13}
& Transport layer
& \covnone & \covfull & \covnone & \covpart
& \covfull & \covpart & \covfull & \covnone
& \covpart & \covnone & \covfull
\\
\midrule

\multirow{3}{*}{
    \makecell[c]{\textbf{AI for THz Service}}}
& \makecell[c]{Edge computing}
& \covnone & \covnone & \covnone & \covnone
& \covfull & \covfull & \covfull & \covnone
& \covfull & \covnone & \covfull
\\
\cline{2-13}
& \makecell[c]{Sensing}
& \covnone & \covnone & \covpart & \covpart
& \covfull & \covfull & \covpart & \covpart
& \covpart & \covnone & \covfull
\\
\cline{2-13}
& \makecell[c]{ISCC}
& \covnone & \covnone & \covnone & \covnone
& \covpart & \covfull & \covnone & \covnone
& \covnone & \covnone & \covfull
\\
\midrule

\multirow{4}{*}{
    \makecell[c]{\textbf{THz for AI}}}
& AI training
& \covpart & \covnone & \covnone & \covpart
& \covfull & \covfull & \covfull & \covpart
& \covpart & \covpart & \covfull
\\
\cline{2-13}
& AI inference
& \covpart & \covnone & \covnone & \covnone
& \covfull & \covfull & \covfull & \covnone
& \covnone & \covnone & \covfull
\\
\cline{2-13}
& Data collection
& \covpart & \covnone & \covnone & \covnone
& \covfull & \covfull & \covpart & \covpart
& \covnone & \covpart & \covfull
\\
\cline{2-13}
& \makecell[c]{THz physical computing}
& \covnone & \covnone & \covnone & \covnone
& \covnone & \covnone & \covnone & \covnone
& \covnone & \covnone & \covfull
\\

\bottomrule
\end{tabularx}

\vspace{3pt}
{\footnotesize
\covfull: substantive coverage;\quad
\covpart: partial or prospective coverage;\quad
\covnone: no substantive coverage identified.
\par}
\vspace{-3mm}
\end{table*}

Meanwhile, the emergence of increasingly complex AI technologies, including large AI models and embodied intelligence, has also revealed substantial potential for THz networks to serve as a key infrastructure for future AI systems.
Specifically, in large-scale AI data centers, the increasing collective communication traffic can become a major system bottleneck, while fixed wired interconnects offer limited flexibility in adapting to evolving workloads. High-capacity and reconfigurable THz links can complement conventional fabrics by providing on-demand connectivity among computing resources, thereby reducing communication latency and allowing network topology to better adapt to AI workloads~\cite{han2025wires}. 
Beyond data centers, THz networks can also interconnect distributed computing and memory resources across terminals, edge servers, and cloud platforms. By supporting dynamic workload placement and seamless coordination among heterogeneous computing nodes, such connectivity enables AI services to exploit distributed resources more flexibly and efficiently.
Moreover, high-resolution THz sensing can provide fine-grained physical-world information for multi-modal AI, supporting embodied agents in environmental perception, understanding, and interaction~\cite{luo2026lowaltitudeembodied}.
At a more fundamental level, conventional AI computing incurs substantial latency and energy overhead from repeated arithmetic operations and memory movement. THz EM structures provide an alternative physical computing paradigm, where learned transformations can be performed directly through wave propagation, potentially reducing digital computation and data movement overhead~\cite{lin2018all}.
Ultimately, as illustrated in Fig. \ref{fig_aithz}, the profound symbiosis of AI-driven methodologies and THz networks marks a transformative paradigm shift to foster innovative solutions for future intelligent wireless systems.

\subsection{Related Work}
Recent studies have reviewed research on applying AI to 6G communication networks, as summarized in Table \ref{tab_ref_survey}.
Specifically, in \cite{Sharma2025Overview}, the authors analyze the specific advantages of multiple key AI optimization techniques in 6G networks, while demonstrating their synergistic effects in enhancing 6G network performance and improving energy efficiency.
From a physical layer perspective, the authors in \cite{zhang2021deep} have investigated deployment strategies for DL, and envision a transformative evolution from conventional block-based designs toward comprehensive end-to-end architectures. 
In addition to offering a comprehensive review of AI applications in the physical layer, the authors of \cite{ye2024artificial} investigate the progress of 3GPP in standardizing AI within this domain.
Beyond the physical layer, AI methodologies have demonstrated significant potential in optimizing higher-layer functions such as intelligent routing, traffic control, and adaptive streaming to ensure end-to-end Quality of Service (QoS) or Quality of Experience (QoE) in heterogeneous environments \cite{tang2021survey}. 
One step further, the work in \cite{cui2025overview} outlines the theoretical foundation for the deep integration of AI and communications, and investigates the bidirectional enabling relationship and key challenges between AI technologies and network technologies in 6G networks. 
As this integration deepens, the concept of AI-native communications \cite{hoydis2020toward} has emerged to fundamentally redefine protocol stacks. Recent practical implementations leveraging digital twins and high-performance computing have proven the industrial viability of AI-native frameworks \cite{cohen2025nvidia}. Furthermore, inspired by the advent of large language models (LLMs), the work in \cite{zhu2025wireless} investigates how wireless foundation models are revolutionizing 6G. By leveraging pre-trained architectures to achieve generalization across diverse communication scenarios, this paradigm is pivotal in driving the evolution toward an AI-native 6G system.

\begin{figure*}[!htbp]
    \centering
    \includegraphics[width=0.8\linewidth]{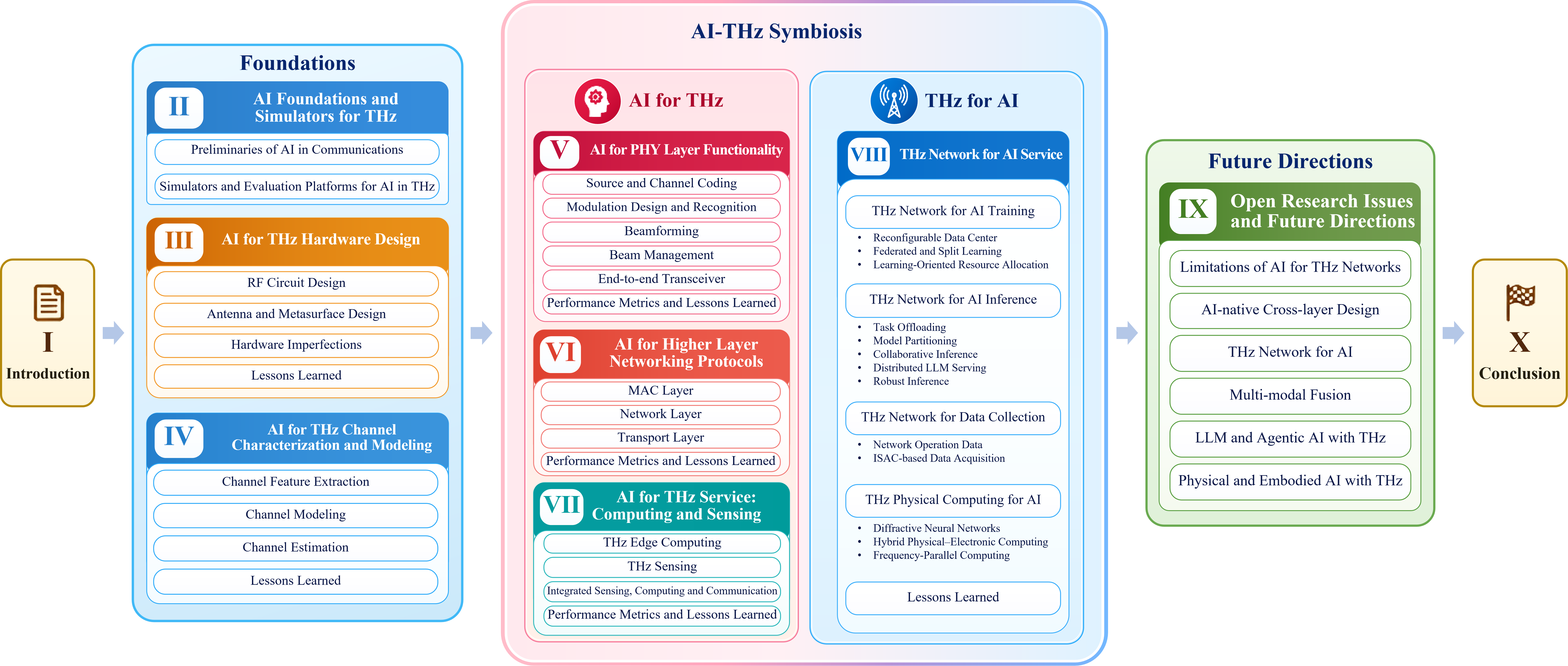}
    \caption{Structure of the survey.}
    \label{fig_structure}
    \vspace{-5mm}
\end{figure*}

Although these surveys comprehensively overview the potential of AI in optimizing various facets of 6G networks and the symbiotic evolution of AI and communications, they often overlook the unique and stringent constraints imposed by the THz band. Moreover, the complex physical properties and highly non-stationary nature of THz signals present challenges in data acquisition and model training \cite{THz_network_challenge}. 
Some specialized efforts have attempted to bridge this gap. For instance, the work in \cite{khan2023machine} provides a review of existing machine learning (ML) based THz beam management techniques, with a particular emphasis on the key characteristics of optimal beam management frameworks. Additionally, the study in \cite{yu2024ai} concentrates on a systematic roadmap tailored primarily for AI-driven beamforming and beam management within THz UM-MIMO systems. However, these works do not encompass the broader network orchestration. 
In contrast, the work in \cite{boulogeorgos2021machine} explores traditional ML applications across various protocol layers, including signal processing, beamforming, and resource management. 
The review in \cite{farhad2023terahertz} highlights the general potential and drawbacks of DL and RL in THz networks.
While these works present a more comprehensive survey of AI techniques for optimizing THz networks, they still lack an in-depth discussion on THz unique issues, like cross far-field and near-field effects, hardware imperfections, and accurate channel modeling. In addition, they have not explored cutting-edge AI technologies for THz.

\vspace{-3mm}
\subsection{Contributions and Organization}
This survey provides a timely and comprehensive survey of the symbiosis of AI and THz paradigms, e.g., AI for THz and THz for AI. 
By identifying the intrinsic challenges across diverse architectural layers, we extensively analyze how various AI paradigms can be leveraged to mitigate specific challenges and optimize network performance in the THz band. 
Beyond investigating how AI can address the intrinsic challenges of THz communications, we further explore the reverse and potentially more transformative question: What is THz ultimately for? After examining the emerging applications that may truly demand the unique capabilities of the THz band, we envision a paradigm shift to \textit{THz for AI}. The next generation of physical AI systems, including humanoid robots, autonomous machines, and embodied agents, will require not only Tbps wireless connectivity for massive multi-modal data exchange, but also high-resolution sensing for perceiving and interacting with the physical world. Thus, THz technology can provide a critical wireless infrastructure bridging AI intelligence in the digital domain with autonomous actions in the physical world. In this sense, instead of being merely another spectrum extension for 6G or beyond, THz is an enabling infrastructure for Physical AI. The main contributions are summarized as follows.

\begin{itemize}
    \item \textbf{AI for THz: A cross-layer perspective.} We provide a comprehensive cross-layer review of how AI can address the fundamental challenges of THz networks, spanning hardware and antenna design, channel modeling and estimation, physical-layer transmission, higher-layer networking, and emerging communication, sensing, and computing services. We also summarize the AI foundations, simulators, and evaluation platforms that support AI-enabled THz research.
    \item \textbf{THz for AI: Enabling next-generation AI infrastructure.} We first examine how THz can serve as a key wireless infrastructure for AI training and inference. Then, we discuss how THz sensing can provide physical-world data for embodied and physical AI. Finally, we review emerging THz physical computing paradigms for efficient AI execution.
    \item \textbf{Toward an AI-THz symbiosis.} Building upon these two complementary directions, we further identify open challenges and future research directions for advancing the symbiosis between AI and THz communications. We highlight key opportunities in AI-native THz networking, THz-enabled AI infrastructure, and emerging paradigms that further deepen their bidirectional integration.
\end{itemize}

The remainder of this survey follows the organization in Fig. \ref{fig_structure}. Sec. \ref{sec_ai} introduces the foundations of AI for THz networks and reviews major simulators and evaluation platforms for THz systems. Then, Sec. \ref{sec_hardware} reviews AI solutions to optimize THz hardware design and mitigate hardware imperfections. In Sec. \ref{sec_channel}, we categorize algorithms to extract channel features and explore generative AI developments to construct precise channel models and achieve accurate estimation. Based on these, Sec. \ref{sec_phy} examines AI applications across core physical layer functionalities. AI methods for higher-layer THz networking design are presented in Sec. \ref{sec_network}. Sec. \ref{sec_service} introduces AI-empowered THz computing and sensing services. Furthermore, we discuss the THz networks for AI training, inference, data collection, and physical computing in Sec. \ref{sec_thz4ai}. Finally, Sec. \ref{sec_future} outlines open challenges and future directions.

\section{AI Foundations and Simulators for THz}
\label{sec_ai}

\subsection{Preliminaries of AI in Communications}
AI techniques for wireless communications can be broadly characterized by how they learn representations and make decisions. Conventional machine learning methods, such as support vector machines and random forests, are suitable for relatively structured problems with engineered features, while deep learning automatically extracts nonlinear and high-dimensional representations from data. Different neural architectures further exploit the structure of wireless information. Convolutional neural networks (CNNs) capture spatial correlations in channel matrices, recurrent neural networks (RNNs) and long short-term memory
networks (LSTMs) model temporal evolution, and graph neural networks (GNNs) represent interactions among users, links, and network nodes. These capabilities have supported channel estimation, beam management, traffic prediction, and resource allocation \cite{b6,zhao2024lstm,he2022gblinks}. 
DRL learns adaptive policies directly from interactions with dynamic network environments. Representative algorithms such as DDPG and PPO have been widely applied to beam control, resource allocation, routing, and other adaptive network functions \cite{ju2022drl,luong2021deep}. 

With advances in generative AI, generative adversarial networks (GANs) and diffusion models can synthesize channel realizations or recover signals from incomplete and noisy observations \cite{zhang2021distributed,diao2026robust}, while Transformer-based models exploit self-attention to capture long-range dependencies for channel estimation, prediction, and beam management \cite{ju2024transformer}. These methods are particularly useful when explicit THz models are difficult to construct, or measurement data are limited. More recently, foundation models and LLMs extend AI from task-specific learning toward reusable representations, reasoning, and cross-task knowledge transfer. In wireless networks, they have been explored for jointly supporting multiple physical-layer functions and higher-layer network management \cite{liu2025llm4wm,wu2024netllm}. This capability is particularly attractive for THz systems, where channel conditions, beam management, and resource allocation are strongly coupled across network functions. Agentic AI further introduces planning, memory, and tool use, allowing an AI agent to interpret network observations, decompose management objectives, and invoke specialized predictors, simulators, or optimization tools \cite{jiang2026large,abou2025agentic}. These developments enable THz systems to manage the complex and dynamic network automatically.

\vspace{-3mm}
\subsection{Simulators and Evaluation Platforms for AI in THz}
This subsection reviews the main simulators and evaluation platforms for AI-enabled THz research, including EM and circuit solvers, channel and ray-tracing tools, network simulators, AI development environments, and measured datasets. At the hardware level, CST provides full-wave EM simulation for antennas, arrays, and other EM structures, while ADS supports RF circuit and system-level simulation. In AI-assisted hardware design, these tools mainly generate physics-based training data or evaluate candidate designs produced by learning and optimization algorithms. Additionally, channel and propagation simulators provide another major source of training and evaluation data. QuaDRiGa \cite{quadriga} generates geometry-based stochastic channels with spatial consistency and temporal evolution, whereas Wireless InSite \cite{remcomwirelessinsite2026} performs site-specific three-dimensional ray tracing based on the propagation environment and antenna configuration. More THz-oriented modeling is provided by TeraMIMO \cite{tarboush2021teramimo}, which incorporates wideband UM-MIMO effects including molecular absorption, spherical-wave propagation, beam split, and beam misalignment. NYUSIM \cite{poddar2024nyusim} instead generates measurement-based statistical channels over its documented 0.5--150~GHz range. 
Moving to the network level, ns-3 is widely used for packet-level simulation of network protocols, traffic, mobility, and resource management. TeraSim further extends ns-3 with THz channel, directional-antenna, PHY, and MAC models, enabling protocol evaluation under THz-specific propagation and highly directional links \cite{hossain2018terasim}. MATLAB-based platforms, particularly Wireless Network Toolbox, also support configurable network nodes, traffic, mobility, custom channel models, and packet-level evaluation, and can be integrated with THz-specific physical-layer models when needed \cite{mathworkswireless2026}.

AI algorithms are commonly implemented in Python using frameworks such as PyTorch and TensorFlow, which provide flexible support for various learning models \cite{paszke2019pytorch,abadi2016tensorflow}. 
More broadly, incorporating native AI interfaces into these simulators could facilitate the development and reproducible evaluation of learning-based THz systems by enabling tighter integration between simulation models and AI algorithms. Sionna represents such a platform by combining differentiable communication models with a Python-based learning environment, while Sionna RT further incorporates differentiable ray-tracing-based propagation modeling \cite{hoydis2022sionna,hoydis2023sionnart}. Beyond these simulators and platforms, measured datasets are also important for AI tasks. A public THz channel dataset \cite{twcdataset2026} covers  indoor, outdoor, near-field, in-vehicle, data-center, and long-range scenarios, with center frequencies up to 400~GHz and more than 600,000 channel impulse responses. Such datasets can complement simulator-generated samples for training, validation, transfer learning, and generalization tests.

\section{AI for THz Hardware Design}
\label{sec_hardware}

\begin{table*}[t]
\centering
\caption{Summary of AI Methods for THz Hardware Design}
\label{tab_hardware_design}
\renewcommand{\arraystretch}{1.05}
\setlength{\tabcolsep}{3pt}
\footnotesize
\begin{tabularx}{\textwidth}{
|>{\centering\arraybackslash}m{2cm}
|>{\centering\arraybackslash}m{2.05cm}
|>{\centering\arraybackslash}m{3.2cm}
|>{\centering\arraybackslash}X
|>{\centering\arraybackslash}m{1.05cm}|}
\hline
\textbf{Design Target}
& \textbf{AI Task}
& \textbf{AI Method}
& \textbf{Design Objective}
& \textbf{Ref.} \\ \hline

\multirow{5}{*}{\shortstack{RF Circuit\\Design}}
& Prediction
& ANN $\&$ transfer learning
& D-band transistor and power-match modeling
& \cite{ariaspurdue_tmtt2024} \\ \cline{2-5}

& Prediction
& Deep CNN
& Passive and PA inverse synthesis
& \cite{hardware_jssc2023} \\ \cline{2-5}

& Prediction
& CNN
& Multi-port RF/sub-THz inverse synthesis
& \cite{karahan_natcomm2024} \\ \cline{2-5}

& Optimization
& RL
& Specification-to-layout sub-THz PA synthesis
& \cite{hardware_isscc2025} \\ \cline{2-5}

& Optimization
& PPO
& Specification-to-layout sub-THz LNA synthesis
& \cite{zhou_isscc2026} \\ \hline

\multirow{9}{*}{\shortstack{Antenna and\\Metasurface\\Design}}
& Prediction
& Extra Trees regression
& THz MIMO antenna gain prediction
& \cite{haque2024machine} \\ \cline{2-5}

& Prediction
& Extra Trees regression
& THz MIMO antenna efficiency prediction
& \cite{haque2025predictive} \\ \cline{2-5}

& Prediction
& Ridge regression
& Eight-port THz MIMO antenna gain prediction
& \cite{albawri_ijcs2025} \\ \cline{2-5}

& \shortstack{Optimization $\&$\\Prediction}
& Gaussian process regression
& Wideband THz antenna parameter optimization
& \cite{hardware_antenna_thz} \\ \cline{2-5}

& \shortstack{Optimization $\&$\\Prediction}
& \shortstack{Random forest}
& High-gain THz antenna design
& \cite{reguig2025ai} \\ \cline{2-5}

& Prediction
& Extra Trees regression
& Metasurface beam-response prediction
& \cite{fakharian_mtcomm2025} \\ \cline{2-5}

& Prediction
& DNN
& Tunable THz beam-steering metasurface synthesis
& \cite{dao_motl2023} \\ \cline{2-5}

& Optimization
& ANN $\&$ transfer learning
& On-demand THz metasurface/transmitarray synthesis
& \cite{jue_olt2025} \\ \cline{2-5}

& Prediction
& CNN
& Broadband pixelated-metasurface synthesis
& \cite{teng_olt2023} \\ \hline
\end{tabularx}

\vspace{-3mm}
\end{table*}
The development of THz hardware design is important to unleash the potential of THz technology. However, the design of RF circuits and antennas increasingly involves high-dimensional parameter spaces and computationally intensive EM simulations, while practical transceivers are further affected by nonlinear hardware impairments. 
In this section, we review how AI addresses these challenges from several functional perspectives. Learning-based optimization is employed to explore complex circuit and structural design spaces, surrogate modeling enables rapid prediction of circuit and EM responses, and inverse-design methods translate target specifications into physical structures or layouts. Meanwhile, data-driven models learn nonlinear distortion characteristics to mitigate hardware imperfections. 
Accordingly, the following discussion focuses on AI-assisted hardware design for THz communications, with particular attention to RF circuits, antennas and metasurfaces, and hardware imperfections. The reviewed hardware designs are summarized in Table~\ref{tab_hardware_design}.

\vspace{-3mm}
\subsection{RF Circuit Design}
RF circuits constitute the fundamental physical infrastructure of THz communication systems. Their design requires iterative optimization across active circuits and electromagnetic structures such as matching networks and power combiners. 
As the operating frequency increases, parasitic effects and electromagnetic coupling become increasingly influential, which makes circuit performance more sensitive to both device parameters and physical layout. Accurate evaluation therefore relies heavily on computationally intensive circuit and full-wave EM simulations. Meanwhile, the growing number of design variables enlarges the feasible design space and makes conventional workflows based on predefined circuit templates and expert-guided tuning increasingly inefficient. 
In contrast, AI techniques demonstrate superior exploration capabilities and thereby facilitate the discovery of optimal circuit design schemes. Moreover, these data-driven approaches exhibit strong generalization performance and overcome the limitations of experience-based methods. Through the capability of data fitting, AI methods significantly accelerate the EM performance verification process.

Early studies mainly use AI for circuit parameter optimization. Specifically, the DRL framework in \cite{hardware_date2020} utilizes sparse subsampling to explore the design space and automatically adjust circuit parameters while meeting target specifications. It converges approximately 40 times faster than traditional algorithms and supports transfer across process nodes. 
The authors in \cite{hardware_dac_2020} combine GNNs with RL to solve the cross-process and cross-topology transistor sizing problem. The GNN extracts structural features from circuit graphs, with transistors as vertices and connections as edges, while RL enables knowledge transfer for design reuse across technology nodes and topologies.
Beyond parameter search, learned models reduce the cost of circuit characterization. For D-band power circuits, \cite{ariaspurdue_tmtt2024} trains an artificial neural network (ANN) on vector load-pull measurements at 130--140 GHz to model the nonlinear dependence of large-signal transistor performance on load impedance and input power. It predicts output power, gain, and efficiency and supports transfer across frequencies and device configurations, facilitating the estimation of power-matching conditions for amplifier design.

More recent studies pursue inverse and end-to-end circuit synthesis, where target performance specifications guide the search for physical implementations. In \cite{hardware_jssc2023}, deep CNN-based EM surrogates reduce repeated full-wave simulations and enable the synthesis of manufacturable passive structures co-designed with power amplifiers. The study in \cite{karahan_natcomm2024} extends this approach to arbitrary multi-port RF and sub-THz structures. Its CNN-based EM emulator predicts the scattering and radiative characteristics of candidate geometries, enabling inverse synthesis of matching networks, filters, power dividers, and combiners, as well as co-design with active circuits.
At the circuit level, the study in \cite{hardware_isscc2025} integrates RL with inverse EM design to jointly address architecture discovery, topology selection, and parameter optimization for 30-120 GHz power amplifiers. This direction is extended in \cite{zhou_isscc2026}, which combines PPO with AI-assisted inverse EM design in a specification-to-layout flow for mmWave and sub-THz low-noise amplifiers (LNAs). Experimental validation includes LNAs covering 24-90~GHz and designs with dual resonances around 85 and 150~GHz, achieving competitive gain and noise performance.

\begin{figure*}[t!]
    \centering
    \includegraphics[width=0.8\linewidth]{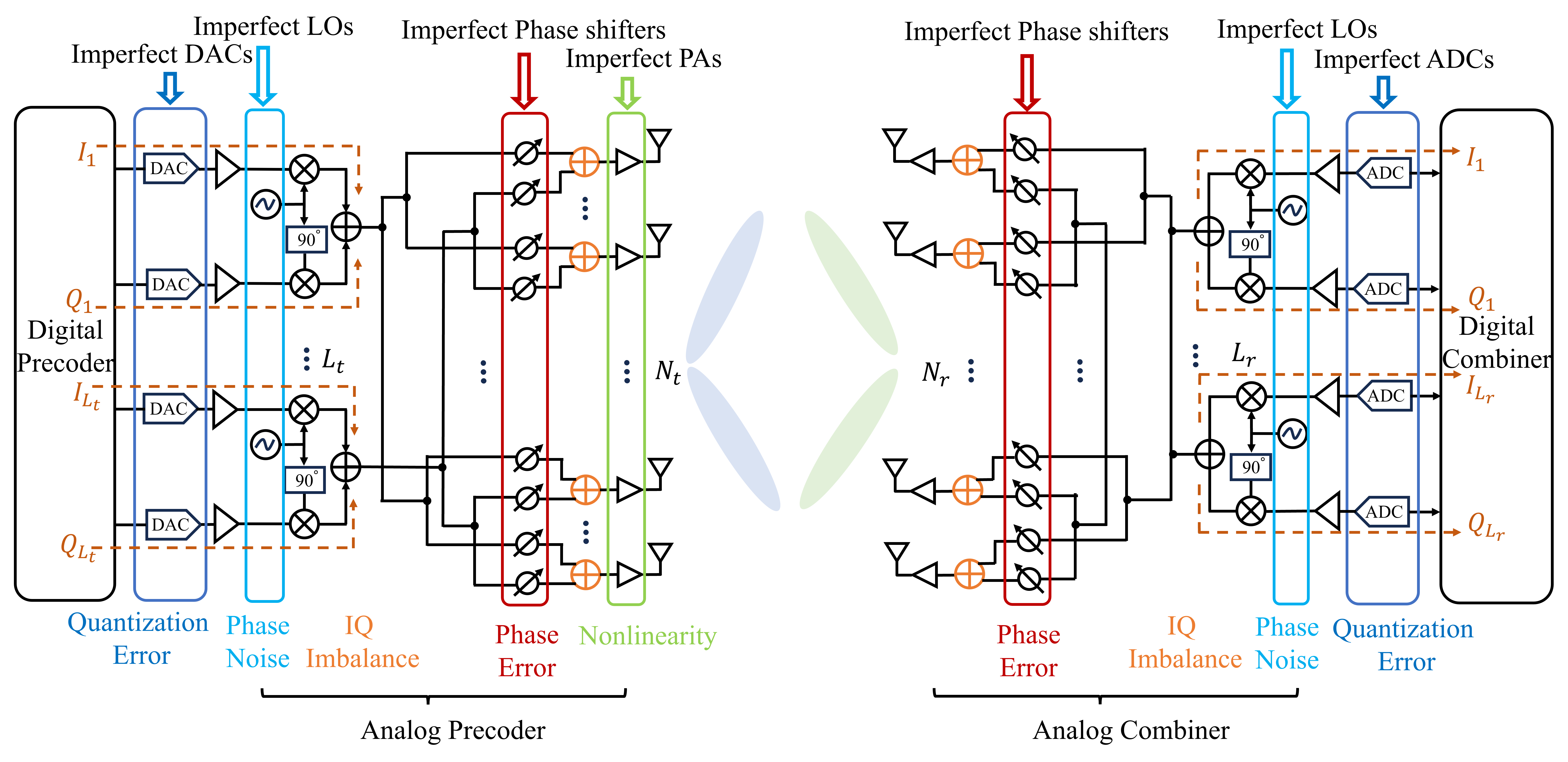}
    \caption{Hardware imperfections in THz UM-MIMO systems~\cite{hardware_imperfection}.}
    \label{fig_hardware_imperfection}
    \vspace{-3mm}
\end{figure*}

\vspace{-3mm}
\subsection{Antenna and Metasurface Design}
The antenna serves as a crucial component of the RF front end in communication systems. Since the THz band suffers from high path loss, stringent requirements are imposed on antenna gain. Simultaneously, the wideband nature of THz systems demands antennas with sufficient bandwidth. Consequently, the optimization of antenna design parameters to balance bandwidth, gain, and size constitutes a significant challenge. 
However, traditional methods often rely on expert knowledge and struggle to efficiently identify optimal solutions within the vast parameter space. Furthermore, the sub-millimeter wavelength of THz signals mandates extremely fine mesh generation during full-wave EM simulations. This avoids severe resonant frequency shifts caused by minute errors but inevitably leads to prohibitive computational resource consumption and extended simulation time.
Recently, metasurface-enabled antennas have received increasing research attention for THz communications ~\cite{zang2021metasurfaces, li2023broadband}. Compared with conventional antennas, metasurfaces composed of subwavelength elements provide additional degrees of freedom for manipulating the amplitude, phase, and polarization of THz waves, thereby facilitating high-gain radiation and flexible beam control. The short wavelength of THz signals also enables the dense integration of a large number of elements within a compact aperture for fine-grained wavefront manipulation. Nevertheless, these additional structural and configurable degrees of freedom substantially enlarge the design space, while strong EM coupling and repeated full-wave evaluation also complicate the optimization process.

To address these complexities, some research explores AI solutions for high-dimensional antenna parameter optimization \cite{hardware_antenna_tap,hardware_antenna_design_magaine,hardware_antenna_review}. Additionally, AI models effectively predict the EM characteristics of antennas and arrays to accelerate the design process. Although initially developed for general frequency bands, these AI-driven design schemes find extensive application in individual antennas \cite{hardware_antenna_tap_r5}, antenna arrays \cite{hardware_antenna_tap_r6,hardware_antenna_tap_r7,hardware_antenna_tap_r8}, and artificial EM media such as metasurfaces \cite{hardware_antenna_tap_r9}. Their ability to improve design efficiency proves particularly valuable for the demanding specifications of THz components.


When operating in the THz band, a fundamental application of AI is fast performance prediction. The studies in \cite{haque2024machine,haque2025predictive} use regression models, including decision trees, random forests, and extra trees, to predict THz MIMO antenna gain from structural parameters. Similarly, the work in \cite{albawri_ijcs2025} uses supervised regression, particularly ridge regression, to predict characteristics such as realized gain for an eight-port three-dimensional THz MIMO antenna. These surrogates learn nonlinear mappings between antenna structures and EM responses from simulation data, reducing full-wave evaluations during design exploration. Beyond prediction, AI surrogates support automatic parameter optimization. In \cite{hardware_antenna_thz}, Gaussian process regression predicts return loss and impedance to guide differential evolution in antenna parameter selection. The study in \cite{reguig2025ai} combines random forest prediction with genetic algorithms and particle swarm optimization for multi-parameter design of high-gain THz antennas. Embedding these surrogates in optimization loops allows more candidate configurations to be evaluated with fewer full-wave simulations.

Metasurface-enabled THz antennas further extend AI-assisted design to higher-dimensional structural and configuration spaces. 
In \cite{fakharian_mtcomm2025}, regression models learn mappings from metasurface coding configurations to far-field responses, enabling rapid performance evaluation. Moving from forward prediction to inverse design, the DNN in \cite{dao_motl2023} maps desired transmission responses to metasurface geometrical parameters for tunable THz beam steering. The study in \cite{jue_olt2025} combines transfer-learning-based ANNs with genetic optimization to reuse EM knowledge across related metasurface design tasks, demonstrating a 2 THz transmitarray with 29.6 dBi simulated gain and 52\% aperture efficiency. Beyond optimizing continuous geometrical parameters, \cite{teng_olt2023} combines CNN surrogates with genetic optimization to search binary pixelated THz metasurface patterns according to target broadband amplitude and phase responses, which extends AI-assisted design toward topology-level structural synthesis.

\vspace{-3mm}
\subsection{Hardware Imperfections}
To realize a robust THz communication system, beyond the physical design of hardware facilities such as RF circuits and antennas, the inherent hardware imperfections within the entire system must be fundamentally addressed.
The increasing carrier frequency and massive bandwidth in the THz band significantly exacerbate the impact of hardware imperfections, which are often negligible in lower-frequency regimes. These amplified impairments severely distort the THz signals, thereby substantially increasing the error probability and degrading system performance.

As illustrated in Fig.~\ref{fig_hardware_imperfection}, THz UM-MIMO systems encounter various impairments across the transceiver chain. 
For instance, while the ultra-high data rates of THz systems demand high-precision digital-to-analog converters (DACs) and analog-to-digital converters (ADCs), prohibitive hardware costs often restrict practical implementations to low-resolution components, thereby introducing severe quantization errors.
Furthermore, amplitude and phase mismatches between the in-phase and quadrature branches in RF front-ends cause I/Q imbalance, leading to detrimental image interference. 
Additionally, nonlinearity in power amplifiers (PAs), the inherent instability of local oscillators (LOs), and imperfect phase shifters at extremely high frequencies generate substantial phase noise. 

Collectively, these imperfections lead to signal constellation deformation, where the received symbols corresponding to different constellation points may overlap, and the demodulation accuracy degrades. 
Conventional signal processing methods, such as minimum mean squared error equalization, are typically mathematically intractable and computationally inadequate for disentangling these compounded nonlinear distortions. 
In contrast, AI-driven methodologies exhibit superior capabilities in learning and compensating for highly complex nonlinearities, thereby effectively mitigating I/Q imbalance, quantization errors, and other associated RF impairments. 
In the following, we review recent advancements in utilizing AI approaches to solve hardware constraints.

Specifically, to address hybrid distortions, an extreme learning machine in \cite{jiang2022novel} processes real and imaginary components of the received signal. This approach achieves joint channel estimation and signal detection and recovers active antenna indices and symbols directly from impaired signals without explicit channel state information (CSI). 
Furthermore, several works deepen the integration of receiver functions to handle hybrid distortions. In \cite{huang2022autoencoder}, the authors exploit autoencoders to learn THz channel and hardware properties, which enables symbol recovery and demodulation under unknown channel models. Similarly, a network in \cite{modulation_Lcomm2022} learns the inverse distortion mapping from received signals to output demodulated bits with improved accuracy. Moreover, the work in \cite{oyekola2024data} proposes a DNN module, which maps impaired signals directly to bits and integrates denoising, demodulation, and channel decoding into a single unit. 
However, these studies generally neglect quantization errors. In \cite{modulation_TCOMM2021}, the authors propose a deep feedforward neural network receiver to effectively recover signals under the dual constraints of extreme one-bit quantization and hybrid hardware impairments. Extending this beyond simple links, the study in \cite{hardware_imperfection} utilizes a DNN in UM-MIMO systems to learn the equivalent channel and effectively overcome the impact of THz hardware imperfections.

\vspace{-3mm}
\subsection{Lessons Learned}

\begin{itemize}
\item \textbf{Surrogate modeling and performance mapping.} Learning-based surrogates approximate the mapping from circuit, antenna, and metasurface parameters to EM responses, reducing repeated simulations and accelerating design-space exploration. However, their accuracy depends on sufficient training coverage of target design.

\item \textbf{Optimization and inverse design.} RL and evolutionary optimization can efficiently search high-dimensional parameter spaces, while learned inverse models directly translate desired responses into circuit layouts or antenna geometries. This progression enables optimization across parameters, topologies, and physical structures, although manufacturability and electromagnetic consistency remain essential constraints for practical designs.

\item \textbf{Complete hardware validation.} Existing AI-assisted designs are often validated at the component level or at mmWave and sub-THz frequencies. Extending them to complete THz transceivers requires accounting for parasitics, packaging and interconnect losses, and interactions among RF components. Additionally, at the digital back end, AI-assisted data conversion and signal processing, including learned compensation, low-precision processing, and feature-domain representations, may help reduce the processing burden under Tbps-rate requirements.

\end{itemize}


\section{AI for THz Channel Characterization and Modeling}
\label{sec_channel}
Accurate channel characterization and reliable CSI remain essential prerequisites for THz system design. Conventionally, channel analysis relies on measurements and traditional analysis algorithms. However, channel measurement campaigns are expensive and time-consuming. Furthermore, high-resolution THz models demand massive datasets, complicating data collection. Severe molecular absorption and complex near-field effects also invalidate traditional mathematical models. Concurrently, ultra-massive antenna arrays impose computational bottlenecks and excessive pilot overhead on conventional estimation techniques. To overcome these limitations, generative AI synthesizes data distributions to solve data scarcity and enable robust estimation. In this section, we review AI applications across channel feature extraction, channel modeling, and channel estimation.

\vspace{-3mm}
\subsection{Channel Feature Extraction}
Channel feature extraction serves as a critical step in channel characterization and modeling. This process entails the distillation of raw measurement data into interpretable channel semantics. Specifically, given raw channel measurements, certain properties, such as channel gain, can be calculated directly, while others, like multipath component (MPC) clustering, tracking, and characterization, require estimation or extraction from the data. 
Traditional methods often suffer from high computational complexity when processing large datasets or struggle to exploit inherent relationships within the data. In contrast, AI-based approaches can significantly accelerate data processing while effectively extracting underlying channel characteristics~\cite{channel_surevy_tap_part1}.

AI-based channel feature extraction can be categorized into three problem types: regression, classification, and clustering. First, regression involves predicting continuous-valued outputs based on input variables. AI models can estimate channel parameters such as path loss, angle information, and delay from measurement data~\cite{channel_feature_doa,channel_feature_attenuation}.
Second, classification distinguishes between different channel conditions, such as identifying whether a channel is line-of-sight (LoS) or non-line-of-sight (NLoS) \cite{channel_feature_los}, or determining the communication scenario in which the channel operates \cite{channel_feature_scenario}.
Third, clustering groups data points into clusters, a common task in MPC clustering. Different MPCs are categorized into clusters using various metrics and methods. Shape-based clustering assesses whether the MPC envelope matches a specific distribution~\cite{channel_feature_cluster_shape}. Optimization-based methods frame clustering as an optimization problem with defined objectives~\cite{channel_feature_cluster_optimization1,channel_feature_cluster_optimization2}. Distance-based methods measure MPC similarity using inter-MPC distances~\cite{channel_feature_cluster_distance_yichen,channel_feature_cluster_distance}. Density-based methods identify clusters based on MPC density~\cite{channel_feature_cluster_density1,channel_feature_cluster_density2}. Vision-based methods replace manual inspection with computer vision techniques~\cite{channel_feature_cluster_cv1,channel_feature_cluster_cv2}. Finally, trajectory-based methods analyze MPC evolution over time to determine clusters~\cite{channel_feature_cluster_trajectory1,channel_feature_cluster_trajectory2}.

\subsection{Channel Modeling}
\label{subsec_channel_modeling}
At THz frequencies, traditional channel modeling techniques encounter significant challenges due to the following reasons. 
On the one hand, the distinct propagation attributes of the THz band introduce severe modeling complexities \cite{han2022terahertz}. In particular, frequency-dependent molecular absorption causes non-linear attenuation. Moreover, the extremely short wavelengths induce pronounced diffuse scattering from rough surfaces, significantly degrading signal power. These phenomena result in highly complex and dynamic channel behaviors that are difficult to describe using analytical or deterministic models.
On the other hand, traditional statistical channel modeling relies on massive empirical data to extract accurate statistical parameters. However, acquiring such extensive datasets through practical measurement is extremely time-consuming and cost-prohibitive in the THz band.

\begin{figure}
    \centering
    \includegraphics[width=0.95\linewidth]{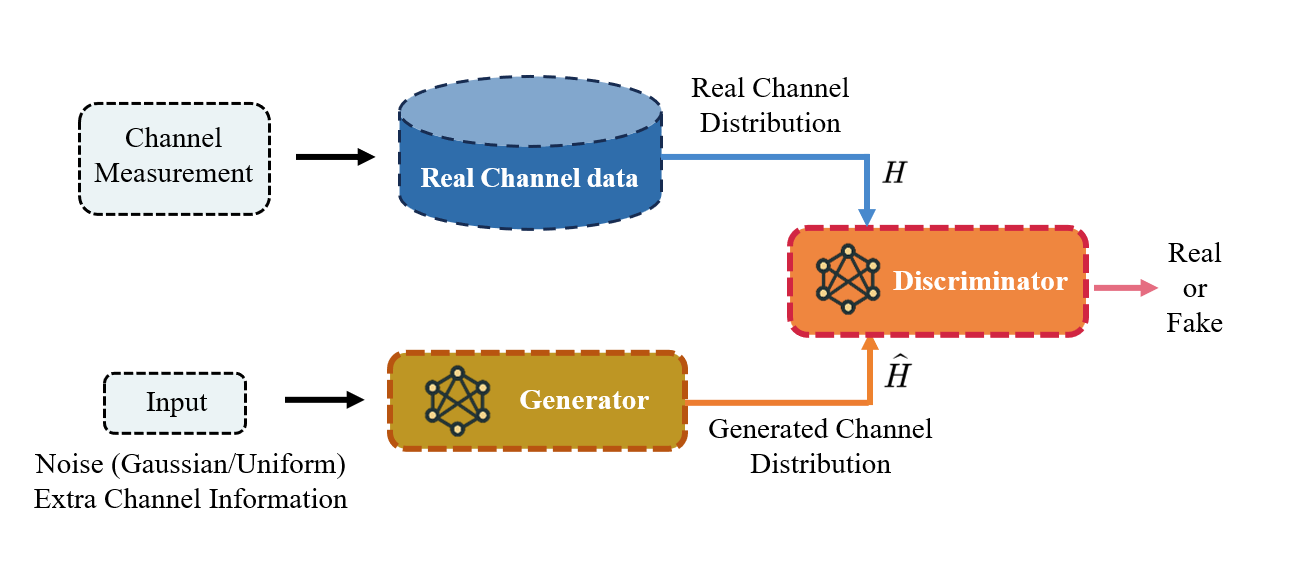}
    \caption{A generic framework for GAN-based channel modeling.}
    \label{fig_gan_channel}
    \vspace{-3mm}
\end{figure}

As a result, AI has emerged as a promising alternative to overcome the limitations of conventional modeling approaches. By leveraging their powerful feature extraction and generalization capabilities, AI-based approaches can capture the complex and dynamic behaviors of channels. 
More specifically, the recent advent of generative AI introduces a novel dimension to this field. Generative models can learn and replicate the underlying distribution of complex datasets, making them well-suited for capturing the inherent stochasticity of wireless channels. Among various DL approaches, GANs have shown considerable potential due to their ability to learn high-dimensional distributions without relying on explicit statistical assumptions. A generic framework for GAN based channel modeling is presented in Fig. \ref{fig_gan_channel}.
Building on this architectural paradigm, the authors in \cite{appro} proposed a conditional variational GAN framework, which introduces a variational layer within the generator architecture to capture the randomness of the wireless channel. By treating the channel modeling task as learning a conditional probability distribution, this method successfully approximates the true channel distribution from limited observational data. The work in~\cite{gan-survey} proposes a GAN-based modeling framework and demonstrates its effectiveness in additive white Gaussian noise channels. The approach in~\cite{channel_gan} introduces a GAN architecture designed to generate synthetic channel matrix samples that closely match the empirical distributions obtained from the clustered delay line channel model. In~\cite{distribution}, a model-driven GAN framework is developed for channel modeling in reconfigurable intelligent surfaces (RIS)-assisted communication systems, illustrating the integration of domain knowledge with data-driven learning. 

While the aforementioned works represent significant progress, they are primarily limited to simplified scenarios and do not fully address the modeling of more complex and practical channel environments.
To address these limitations, a Transformer-based GAN architecture has been proposed to model multipath components by generating parameters such as delay, angle, and power coefficients for each path~\cite{hu2024transfer}. The Transformer structure enhances the model’s ability to capture structural dependencies among multipath components, providing a richer and more expressive representation of the channel. Furthermore, the proposed method incorporates transfer learning strategies, enabling effective adaptation to data-scarce scenarios. This approach significantly improves the generalization capability and practical applicability of GAN-based channel modeling techniques.

\begin{figure}[t!]
    \centering
    \includegraphics[width=0.95\linewidth]{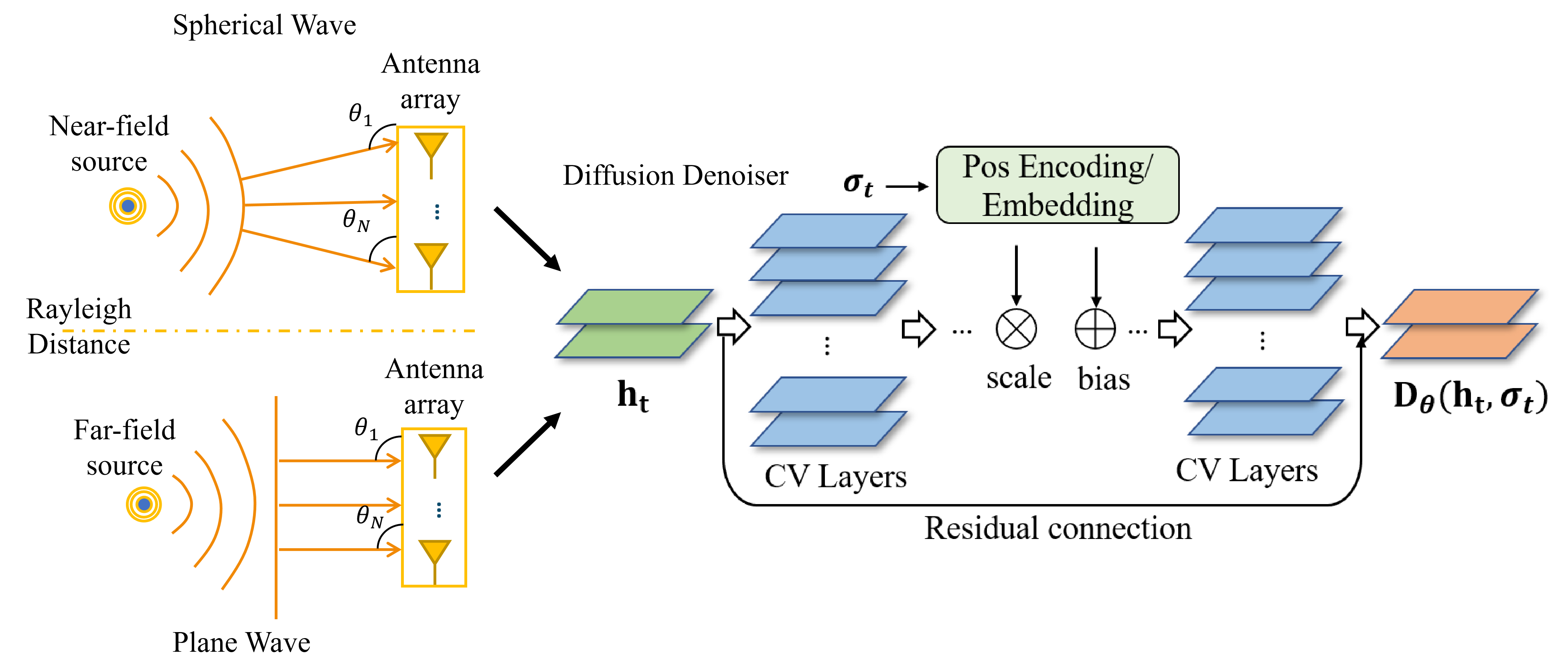}
    \caption{Cross-field channel estimation based on a diffusion model \cite{hu2025diffpace}.}
    \label{fig_diffusion_ce}
    \vspace{-3mm}
\end{figure}

\subsection{Channel Estimation}
Subsequent to the theoretical characterization and modeling of the propagation environment, channel estimation (CE) constitutes a fundamental prerequisite for establishing reliable communication links. It directly dictates the performance of successive functionalities, including precoding, signal detection, and resource allocation. To compensate for severe propagation losses, THz systems widely adopt UM-MIMO systems. However, this massive scale introduces formidable computational complexity. Furthermore, stringent hardware and power constraints compel the use of hybrid beamforming architectures \cite{han2021hybrid}. This specific design relies on limited RF chains to drive massive antennas. Consequently, a severe dimensionality mismatch emerges between high-dimensional channels and compressed received signals. Traditional methods often rely on prior channel statistics to achieve satisfactory estimation accuracy, yet such statistical information is typically unavailable in practice and difficult to estimate reliably. To overcome these limitations, recent research proposes various AI-driven solutions. These approaches effectively extract inherent physical channel characteristics to achieve superior estimation accuracy under complex hardware constraints.

Broadly, AI-based CE methods can be categorized into two main classes, data-driven approaches and model-driven approaches, as shown in Table~\ref{tab_ai_ce}.
Data-driven methods utilize black-box neural networks trained end-to-end with large datasets without explicit channel models or algorithmic priors. 
For instance, DNNs can estimate doubly selective channels directly \cite{b21} or denoise received signals before conventional estimation \cite{b22}. 
In~\cite{b24}, a deep CNN is trained with extensive measurement data to directly estimate key parameters required for channel reconstruction. Similarly, a spatial-frequency CNN in \cite{b6} exploits both spatial and frequency correlations by processing channel matrices of adjacent subcarriers simultaneously.
Unconstrained by traditional structures, these approaches achieve superior accuracy given well-designed architectures and sufficient data.

Conversely, model-driven methods incorporate domain knowledge into the network design, typically by unfolding traditional iterative algorithms into DNN layers. 
This strategy trains fewer parameters to yield efficient interpretable networks.
A prominent foundation is the Approximate Message Passing (AMP) algorithm.
For example, the work in~\cite{b30} unfolds the AMP algorithm into a DNN, allowing its parameters to be learned and optimized from data. 
In \cite{b17}, the authors replace the traditional shrinkage function in AMP with a denoising CNN, thereby significantly enhancing noise robustness. 
Further extending this line of work, another AMP-based framework proposed in \cite{b15} enhances beamspace CE by integrating a Gaussian mixture shrinkage function to leverage prior structural knowledge.
The integration of DL further enhances the estimation accuracy of AMP-based methods. However, these model-driven approaches struggle to exploit complex features beyond simple sparsity and face performance bottlenecks under low pilot overheads.

To address this, generative AI approaches leverage data-driven generative priors to capture intrinsic channel structures and enable superior estimation accuracy even with limited measurements.
For instance, the framework in \cite{ce_gan_jsac2021} integrates GANs with compressed sensing to learn compact channel representations. It maps received signals to a latent space via iterative optimization to ensure adaptability without explicit channel statistics. Another robust approach is the diffusion model \cite{diffusion_beat_gan}. The study in \cite{ce_dm_twc2023} applies a probabilistic diffusion framework to learn MIMO channel statistics and reconstructs signals accurately through Bayesian sampling.

Although the aforementioned approaches achieve state-of-the-art estimation quality and exceptional generalization capabilities, they mostly focus on far-field scenarios. 
In the near-field region, the conventional planar wave assumption breaks down, and propagation is instead characterized by spherical wavefronts. 
Consequently, channel steering vectors depend on both angular direction and transmission range. This dual dependence drastically increases computational complexity. To mitigate this overhead, the study in \cite{Elbir2023Near} deploys CNNs to map received signals to near-field parameters and integrates federated learning (FL) to reduce data transmission. Furthermore, the work in \cite{hu2025diffpace} introduces DiffPace for hybrid near-field and far-field channel estimation, as shown in Fig. \ref{fig_diffusion_ce}. This diffusion-based framework trains on a hybrid planar-spherical wave model to capture a unified channel distribution. 
\begin{table}[t]
\centering
\caption{Summary of AI-Based Channel Estimation Methods}
\label{tab_ai_ce}

\begingroup
\scriptsize
\renewcommand{\arraystretch}{1.15}
\setlength{\tabcolsep}{1.5pt}

\renewcommand{\tabularxcolumn}[1]{m{#1}}

\begin{tabularx}{\columnwidth}{
|>{\centering\arraybackslash}m{1.20cm}
|>{\centering\arraybackslash}m{0.85cm}
|>{\centering\arraybackslash}m{1.20cm}
|>{\centering\arraybackslash}m{1.50cm}
|>{\centering\arraybackslash}X
|>{\centering\arraybackslash}m{0.55cm}|}
\hline

\textbf{Paradigm} &
\textbf{Field} &
\shortstack{\textbf{AI}\\\textbf{Task}} &
\textbf{AI Method} &
\shortstack{\textbf{Exploited}\\\textbf{Information}} &
\textbf{Ref.} \\ \hline

\multirow{6}{*}{\shortstack{Data-\\driven}}
& \shortstack{Cross-\\field}
& Estimation
& Deep CNN
& Hybrid spherical-/planar-wave structure
& \textbf{\cite{b24}} \\ \cline{2-6}

& Far-field
& Estimation
& \shortstack{CNN}
& Spatial-frequency correlation
& \cite{b6} \\ \cline{2-6}

& Near-field
& Estimation
& CNN $\&$ FL
& Received pilots and angle-range dependence
& \textbf{\cite{Elbir2023Near}} \\ \cline{2-6}

& Far-field
& Estimation
& \shortstack{GAN}
& Learned channel manifold
& \cite{ce_gan_jsac2021} \\ \cline{2-6}

& Far-field
& Estimation
& \shortstack{Diffusion\\model}
& Learned channel distribution
& \cite{ce_dm_twc2023} \\ \cline{2-6}

& \shortstack{Cross-\\field}
& Estimation
& \shortstack{Diffusion\\model}
& Planar-/spherical-wave distribution
& \textbf{\cite{hu2025diffpace}} \\ \hline

\multirow{3}{*}{\shortstack{Model-\\driven}}
& Far-field
& Estimation
& \shortstack{DNN unfolding\\AMP}
& AMP structure and channel sparsity
& \cite{b30} \\ \cline{2-6}

& Far-field
& Estimation
& \shortstack{CNN-enhanced\\AMP}
& AMP structure and learned denoiser
& \cite{b17} \\ \cline{2-6}

& Far-field
& Estimation
& \shortstack{AMP with\\Gaussian-\\mixture}
& Beamspace sparsity and Gaussian-mixture prior
& \cite{b15} \\ \hline

\end{tabularx}
\endgroup
\vspace{-3mm}
\end{table}

\subsection{Lessons Learned}
\begin{itemize}
\item \textbf{Structured channel representation.}
AI can efficiently extract path loss, angular, delay, and multipath features from large measurement datasets, providing structured channel information for subsequent modeling and estimation. 
More integrated frameworks could further connect feature extraction with downstream channel reconstruction to reduce repeated processing and exploit correlations among channel parameters.

\item \textbf{Generative channel modeling and estimation.}
Generative models can learn the distribution of complex THz channels and provide informative priors for channel estimation under limited observations.
Combining physical propagation knowledge with generative learning can further constrain the learned distribution, allowing channel modeling and estimation to reinforce each other through a shared representation of THz channel statistics.

\item \textbf{Cross-field channel learning.}
Most existing AI-based estimators target either far-field or near-field propagation, although THz UM-MIMO systems can operate across both regimes. 
Future models should learn unified cross-field channel representations for consistent modeling and estimation without separate field-specific designs.

\end{itemize}

\section{AI for PHY Layer Functionality}
\label{sec_phy}
Following the channel characterization in Sec. \ref{sec_channel}, efficient physical layer operations become the cornerstone of THz communications. AI inherently excels at signal processing, optimization, and forecasting, which directly resolve the rigorous bottlenecks encountered in THz physical layer designs. Guided by this alignment, this section systematically investigates AI applications across key functionalities. 
We begin by exploring how the non-linear mapping and feature extraction superiority of DL enable highly robust source and channel coding, alongside intelligent modulation design and recognition.
Subsequently, we review how AI resolves prohibitive UM-MIMO complexities to deliver computationally efficient beamforming and beam management. 
Furthermore, we examine how DL methods can facilitate end-to-end transceiver designs. Ultimately, these advancements establish the AI radio access network (AI-RAN). This transformative paradigm natively embeds intelligence into network infrastructures to enable fully autonomous operations. The deep integration of AI within the THz physical layer is indispensable to realizing this overarching vision.

\subsection{Source and Channel Coding}
Physical layer operations fundamentally depend on source coding to compress data and channel coding to correct errors. To further improve transmission efficiency, joint source-channel coding (JSCC) integrates both processes into a cohesive framework. 
Traditionally, coding designs rely on rigorous mathematical derivations for idealized models like the AWGN channel. However, THz propagation environments exhibit complex characteristics and extreme interference fluctuations. Consequently, traditional coding paradigms degrade significantly when real channels deviate from standard models.
To overcome these limitations, data-driven methods offer a robust alternative. AI approaches learn underlying environmental statistics to optimize coding strategies adaptively. This capability ensures reliable JSCC performance even under extremely low signal-to-interference-plus-noise ratios. The following discussion reviews several AI-empowered coding schemes with significant potential for THz communication systems.

For instance, the authors in \cite{jiang2019turbo} proposed the Turbo autoencoder, which achieves near-optimal error correction without predefined channel assumptions. To address low-latency scenarios, the study in \cite{xuan2022low} formulates JSCC as an autoencoder optimization problem. It employs bidirectional LSTM and sinusoidal representation networks to map continuous sources directly to channel symbols. Furthermore, an adaptive-rate JSCC scheme in \cite{dai2022nonlinear} utilizes nonlinear transform networks and an AI-learned conditional entropy model to allocate resources dynamically based on source information density. This enhances overall coding gain and bandwidth efficiency.
In harsh environments, generative networks offer unprecedented robustness for JSCC. The study in \cite{erdemir2023generative} integrates pre-trained generative models to maximize the semantic quality of reconstructed images under degraded conditions. 
Moreover, AWGN and diverse unknown interference frequently corrupt practical signals, particularly in dense THz networks. To tackle this, the framework in \cite{wu2025icdm} exploits diffusion models. It estimates prior probability distributions for target signals and complex interference to solve the maximum a posteriori inference problem at the receiver, thereby realizing high-precision interference-canceling JSCC.

\subsection{Modulation Design and Recognition}
In the THz band, massive bandwidths and transmission rates pose stringent requirements on modulation designs, demanding both high spectral efficiency and low computational overhead.
Simultaneously, receivers face unprecedented challenges in modulation recognition. Distance-dependent molecular absorption induces drastic bandwidth fluctuations during user mobility. This phenomenon severely complicates reliable format identification under signal distortions.
To address these bottlenecks, AI methodologies provide compelling solutions. Data-driven algorithms optimize complex modulation schemes under low-complexity constraints. Furthermore, machine learning models demonstrate superior classification capabilities to achieve more accurate and reliable signal recognition than traditional decision frameworks.

For modulation design, the K-nearest neighbors algorithm in \cite{liu2020machine} assists adaptive index modulation by selecting optimal transmission modes to enhance system throughput. This approach evaluates real-time signal-to-noise ratios to transform link adaptation into a classification problem. Furthermore, the study in \cite{wang2025generation} developed a diffractive autoencoder neural network for THz pulse modulation. This architecture integrates a physical diffraction model to train layer parameters directly. It precisely controls pulse characteristics and circumvents complex manual configurations of diffractive surfaces.

Beyond modulation design, AI enables automatic modulation recognition at the receiver. Under harsh weather conditions or complex multipath scattering environments, THz waveforms experience severe fluctuations, which significantly exacerbate the baseband signal processing burden and reduce overall operational efficiency. 
To address this, the study in \cite{wu2022modulation} evaluates CNN and LSTM architectures. These models learn constellation features and time-series patterns to achieve accurate modulation classification. Moreover, CNNs in \cite{hall2023deep} extract translation-invariant features from in-phase and quadrature samples. Combined with majority voting boosting, this approach performs real-time joint recognition of modulation formats and bandwidths. This capability allows rapid receiver adaptation for robust communication in dynamic channels.

\begin{figure}[t]
\centering
\includegraphics[width=0.95\linewidth]{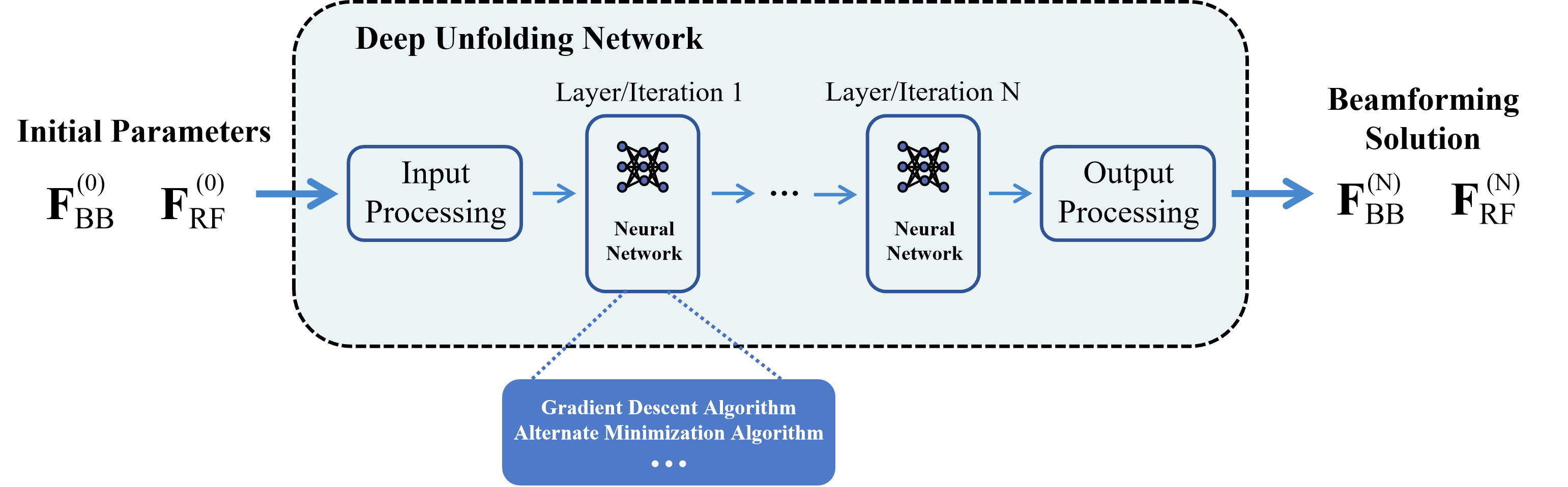}
\caption{Deep unfolding networks for beamforming.}
\label{fig_beam_dun}
\vspace{-3mm}
\end{figure}

\subsection{Beamforming}
\label{Beamforming}
Benefiting from the sub-millimeter wavelength of THz signals, antenna arrays can integrate an ultra-massive number of antenna elements. This facilitates the deployment of the THz UM-MIMO systems mentioned in Sec. \ref{sec_channel}. 
Beamforming technology adjusts antenna phases and amplitudes to achieve coherent signal superposition. This directional gain effectively compensates for severe path loss. However, massive antennas and users generate extremely high-dimensional optimization problems. Traditional iterative algorithms require complex matrix inversions and fail to satisfy real-time low-latency constraints \cite{yu2024ai}.
Furthermore, THz UM-MIMO architectures face unique propagation challenges. Large array apertures extend the Rayleigh distance to introduce complex near-field and mixed-field effects. Simultaneously, wide bandwidths and massive arrays cause signal propagation delays to exceed sampling periods. This temporal mismatch induces the beam squint effect. To overcome these bottlenecks, recent research investigates AI-driven beamforming solutions, which can be broadly classified into model-based and data-driven approaches, as summarized in Table \ref{tab_beam}.

Model-based approaches incorporate domain knowledge to guide beamforming. A leading paradigm is deep unfolding to combine DL with iterative optimization, as illustrated in Fig. \ref{fig_beam_dun}. For instance, the framework in \cite{unfolding_tsp2023} unfolds projected gradient descent to design analog beamformers alongside least squares digital beamformers. Similarly, the network in \cite{BF_unfold3_lwc2022} unfolds alternate minimization iterations to approximate full-digital spectral efficiency with low complexity. To minimize symbol error rates, the method in \cite{BF_unfold2_tvt2022} unfolds a gradient descent algorithm for the non-convex objective function.
Furthermore, the integration of a GAN with unfolding achieves near-optimal designs under minimal CSI feedback \cite{BF_unfold1_twc2021}. However, traditional unfolding terminates prematurely and lacks explicit convergence guarantees. To resolve this, fixed point networks in \cite{yu2025deep} iterate data repeatedly within a single module until stabilization. This mechanism ensures rigorous convergence to yield superior solutions.

In contrast, data-driven approaches learn optimal beamforming solutions directly from data. For instance, autoencoders in \cite{BF_dnn2_tvt2019} and CNNs in \cite{BF_cnn5_jstsp2021} extract channel features to predict hybrid precoding matrices directly. Beyond supervised models, DRL proves highly effective for dynamic optimization. The DDPG algorithm in \cite{drl_beamforming_gc2020} jointly optimizes the digital beamforming at the base station and the analog phase shift matrices of multiple RIS. To mitigate unique THz wideband beam split effects, a hybrid network in \cite{beamforming_suint_letters2023} computes the mean channel covariance to infer precoders through separate neural sub-networks. Under near-field conditions, the study in \cite{fallah2024near} applies RL to configure sub-array phases for precise beam focusing. This framework further uses transfer learning to reduce training overhead and accelerate convergence.

\begin{table*}[t]
\centering
\caption{Summary of AI Methods in Beamforming and Beam Management}
\label{tab_beam}
\renewcommand{\arraystretch}{1.05}
\setlength{\tabcolsep}{2pt}
\scriptsize
\begin{tabularx}{\textwidth}{
|>{\centering\arraybackslash}m{1.4cm}
|>{\centering\arraybackslash}m{1.4cm}
|>{\centering\arraybackslash}m{1.05cm}
|>{\centering\arraybackslash}m{1.3cm}
|>{\centering\arraybackslash}m{2.5cm}
|>{\centering\arraybackslash}X
|>{\centering\arraybackslash}m{2.5cm}
|>{\centering\arraybackslash}m{0.75cm}|}
\hline
\textbf{Task}
& \textbf{Classification}
& \textbf{Field}
& \textbf{AI Task}
& \textbf{AI Method}
& \textbf{THz-Specific Focus / Objective}
& \textbf{Metric}
& \textbf{Ref.} \\ \hline

\multirow{10}{*}{Beamforming}
& \multirow{5}{*}{Model-based}
& Far-field
& Optimization
& Deep unfolding
& THz hybrid precoding under massive-array constraints
& Spectral efficiency (SE)
& \textbf{\cite{unfolding_tsp2023}} \\ \cline{3-8}

& & Far-field
& Optimization
& Deep unfolding
& Low-complexity hybrid precoding
& SE
& \cite{BF_unfold3_lwc2022} \\ \cline{3-8}

& & Far-field
& Optimization
& Deep unfolding
& Minimum-SER hybrid transceiver design
& Symbol error rate (SER)
& \cite{BF_unfold2_tvt2022} \\ \cline{3-8}

& & Far-field
& Optimization
& Deep unfolding
& Hybrid precoding with reduced CSI feedback
& SE
& \cite{BF_unfold1_twc2021} \\ \cline{3-8}

& & Near-field
& Optimization
& Fixed point network
& Near-field beam focusing for XL-MIMO
& SE
& \textbf{\cite{yu2025deep}} \\ \cline{2-8}

& \multirow{5}{*}{Data-driven}
& Far-field
& Optimization
& Autoencoder
& Low-complexity hybrid precoding
& SER $\&$ SE
& \cite{BF_dnn2_tvt2019} \\ \cline{3-8}

& & Far-field
& Optimization
& CNN
& THz joint radar-communications hybrid beamforming
& SE
& \textbf{\cite{BF_cnn5_jstsp2021}} \\ \cline{3-8}

& & Far-field
& Optimization
& DRL
& Multi-hop RIS-assisted THz coverage extension
& Coverage
& \textbf{\cite{drl_beamforming_gc2020}} \\ \cline{3-8}

& & Far-field
& Optimization
& DNN
& Wideband THz beam-squint mitigation
& Sum rate
& \textbf{\cite{beamforming_suint_letters2023}} \\ \cline{3-8}

& & Near-field
& Optimization
& DRL $\&$ transfer learning
& CSI-independent near-field spot beam focusing
& Convergence speed
& \textbf{\cite{fallah2024near}} \\ \hline

\multirow{13}{*}{\shortstack{Beam\\Management}}
& \multirow{7}{*}{\shortstack{Beam\\Alignment}}
& Far-field
& Classification
& Random forest
& Low-complexity THz analog beam selection
& Sum rate
& \cite{ma2019low} \\ \cline{3-8}

& & Far-field
& Classification
& DNN
& Low-overhead hierarchical alignment
& Alignment accuracy
& \cite{yang2023hierarchical} \\ \cline{3-8}

& & Far-field
& Prediction
& DNN
& Interactive AoA alignment with adaptive probing
& Beamwidth
& \textbf{\cite{khalili2021single}} \\ \cline{3-8}

& & Far-field
& Optimization
& DRL
& Environment-adaptive mmWave/THz codebooks
& Beamforming gain
& \textbf{\cite{zhang2021reinforcement}} \\ \cline{3-8}

& & Far-field
& Prediction
& DNN
& Grid-free beam synthesis
& SNR
& \cite{heng2023grid} \\ \cline{3-8}

& & Far-field
& Prediction
& Deep unfolding
& Beam-split-assisted wideband AoA estimation
& Alignment accuracy
& \textbf{\cite{chen2025exploiting}} \\ \cline{3-8}

& & Near-field
& Optimization
& Signal model inspired online learning
& CSI-free wideband THz beam focusing
& Beamforming gain
& \textbf{\cite{zhang2023deep}} \\ \cline{2-8}

& \multirow{4}{*}{\shortstack{Beam\\Tracking}}
& Far-field
& Prediction
& LSTM
& Mobility-aware beam tracking
& Tracking error
& \cite{lim2021deep} \\ \cline{3-8}

& & Far-field
& Prediction
& Ensemble learning
& THz beamspace channel tracking and beam selection
& Tracking error
& \textbf{\cite{zarini2023intelligent}} \\ \cline{3-8}

& & Far-field
& Optimization
& DRL
& Adaptive V2V beam switching and resource allocation
& Throughput
& \textbf{\cite{ahmed2023deep}} \\ \cline{3-8}

& & Near-field
& Optimization
& DRL
& Mobility-model-free near-field THz beam tracking
& Beamforming gain
& \textbf{\cite{park2024robust}} \\ \cline{2-8}

& \multirow{2}{*}{\shortstack{Beam\\Prediction}}
& Far-field
& Prediction
& RNN $\&$ meta-learning
& Sample-efficient mmWave/THz blockage prediction
& Accuracy
& \textbf{\cite{kalor2021prediction}} \\ \cline{3-8}

& & Far-field
& Classification
& Multi-modal
& Vision-position beam prediction
& Accuracy
& \textbf{\cite{charan2022vision}} \\ \hline

\end{tabularx}
\vspace{-3mm}
\end{table*}

\subsection{Beam Management}
Although beamforming mitigates severe path loss, the requirement of real-time CSI in THz UM-MIMO entails prohibitive pilot overhead. Therefore, beam management serves as an essential mechanism to establish and maintain communication links without complete channel knowledge. This mechanism comprises beam alignment, tracking, and prediction. Specifically, beam alignment identifies the optimal beam pair. Beam tracking maintains connections during user mobility. Furthermore, beam prediction proactively determines beam directions to minimize search overhead. However, narrow THz beams render traditional exhaustive scanning impractical for rapid alignment. Simultaneously, the high directionality leaves links vulnerable to outages and necessitates highly efficient tracking schemes \cite{attaoui2022initial}. Recent research leverages DL and RL approaches to tackle these challenges. The following discussion details these AI-driven methodologies and summarizes them in Table \ref{tab_beam}. Although some included studies focus on the mmWave band, their architectures are highly extensible to THz systems. For clarity, references specifically tailored to the THz band are highlighted in bold.

To address the complexity of beam alignment, the study in \cite{ma2019low} formulates a classification problem and utilizes random forests to predict optimal beam indices. To minimize exhaustive search latency, the framework in \cite{yang2023hierarchical} jointly trains probing codebooks and predictors for rapid hierarchical alignment. Furthermore, an interactive DL method in \cite{khalili2021single} progressively adjusts beamwidths and directions to overcome traditional single-scan inaccuracies. Addressing unique THz propagation, the authors in \cite{chen2025exploiting} exploit beam split effects through deep unfolding for wideband alignment. Similarly, an online learning framework in \cite{zhang2023deep} achieves near-field beam focusing. However, these works primarily rely on fixed grid-based codebooks. To overcome this limitation, an RL scheme in \cite{zhang2021reinforcement} adaptively adjusts codebooks according to environmental dynamics. Going a step further to eliminate the grid constraint, a site-specific DL pipeline in \cite{heng2023grid} synthesizes continuous transmit and receive beams from minimal measurements to achieve grid-free high-precision alignment.

Once links are established, beam tracking ensures stability. Some research exploits predictive models for robust tracking. For example, LSTM networks in \cite{lim2021deep} predict future channel distributions to enable adaptive beam control in high-mobility scenarios. Ensemble learning techniques in \cite{zarini2023intelligent} capture the temporal evolution of THz channels. Alternatively, RL demonstrates distinct advantages in dynamic environments. The framework in \cite{ahmed2023deep} leverages DRL to autonomously dictate beam switching and resource allocation without explicit channel models. Extending this capability, the study in \cite{park2024robust} applies similar strategies to solve robust tracking problems under near-field THz propagation conditions.

Regarding beam prediction, the work in \cite{kalor2021prediction} combines RNNs with meta-learning to anticipate THz link blockages. This methodology optimizes initialization parameters from diverse deployments to achieve accurate predictions in new environments with minimal samples. Furthermore, multi-modal techniques introduce an innovative predictive dimension. The study in \cite{charan2022vision} fuses positional and visual information, where a DNN processes these concatenated features to realize rapid and high-precision beam prediction.

\subsection{End-to-end Transceiver}
In practical communication scenarios, the complex and dynamic THz channels pose significant challenges to physical layer design. Although the previously discussed physical functionalities improve specific modules, they essentially substitute only isolated blocks of the communication system. Consequently, such block-level optimization approaches often fail to achieve the global optimum for the entire link. To address this limitation, AI facilitates the optimization of the communication system in an end-to-end manner. Specifically, by treating the entire communication process from the transmitter to the receiver as a black box, DL models can directly learn and fit the signal transformation. This paradigm enables the realization of end-to-end communication performance optimization without the requirement for explicit channel information or the modeling of hardware imperfections \cite{e2e_dl_for_phy,e2e_demon,e2e_over_air,e2e_overair_spawc,e2e_rl,e2e_appro,e2e_liye_agnostic,e2e_additive}.

Recent literature explores various end-to-end transceiver designs for the THz band. For instance, the work in \cite{zhang2021deep} provides a comprehensive review of the utilization of end-to-end DL methods such as autoencoders and model-driven deep unfolding, which address the global joint optimization of transceiver design. Additionally, the study in \cite{marasinghe2025phase} proposes a CNN-based receiver architecture that coordinates with trainable constellations and pilot schemes to solve the end-to-end optimization problem for THz point-to-point high data rate links.
To further enhance the feasibility of practical deployment, some works focus on the reduction of model complexity. The work in \cite{zhao2025transfer} exploits transfer learning techniques to migrate feature knowledge from transceivers built with Transformer and LSTM networks to lightweight models in target scenarios. This strategy effectively addresses the digital pre-distortion problem in THz transceivers and achieves performance optimization under low-complexity requirements. Furthermore, the study in \cite{che2022low} adopts binary neural networks where weights and activations are constrained to binary values to reduce computational complexity at high throughput rates. Combined with generative adversarial networks, this approach realizes the joint training and optimization of the transmitter and receiver in an end-to-end manner.

\subsection{Performance Metrics and Lessons Learned}
For AI-enabled THz physical-layer functions, the commonly used performance metrics mainly reflect transmission efficiency and the effectiveness of beam establishment and maintenance.
Beamforming is primarily evaluated by spectral efficiency and symbol error rate, while CSI feedback overhead and communication coverage are also considered in specific designs.
For beam management, beam alignment focuses on alignment accuracy together with computational complexity, whereas beam tracking and prediction emphasize tracking robustness and prediction accuracy.
The reviewed PHY studies suggest the following design insights.
\begin{itemize}
\item \textbf{Model-data complementarity}.
Model-based methods embed optimization structures and physical knowledge into learning, providing structured solutions but potentially inheriting limitations from the underlying iterative procedures.
Data-driven methods directly learn beamforming or beam-control mappings and offer flexibility in complex THz environments, although their performance relies more strongly on representative training data.
How to effectively combine these two paradigms remains a research direction, intending to exploit model-based structural priors while preserving data-driven flexibility.

\item \textbf{Cross-field beam management}.
Existing beam alignment and tracking methods are mostly designed separately for far-field or near-field propagation, while THz UM-MIMO links may transition between these regimes.
Future schemes should jointly support angular alignment in the far field and range-angle beam focusing in the near field, enabling consistent beam acquisition and maintenance across mixed-field conditions.

\end{itemize}
\section{AI for Higher Layer Networking Protocols}
\label{sec_network}
While preceding physical layer discussions establish the AI-RAN foundation, higher-layer networking ensures robust end-to-end performance. However, THz networks exhibit extreme dynamism, heterogeneous architectures, and diverse service requirements. Traditional methods prove inadequate for managing these intricacies. Conversely, AI empowers networks with essential adaptability and real-time processing. Integrating intelligence into higher-layer protocols extends the AI-RAN framework to significantly enhance network management efficiency. Accordingly, this section investigates AI applications across the THz MAC layer, network layer, and transport layer.

\begin{figure}[t!]
    \centering
    \includegraphics[width=0.95\linewidth]{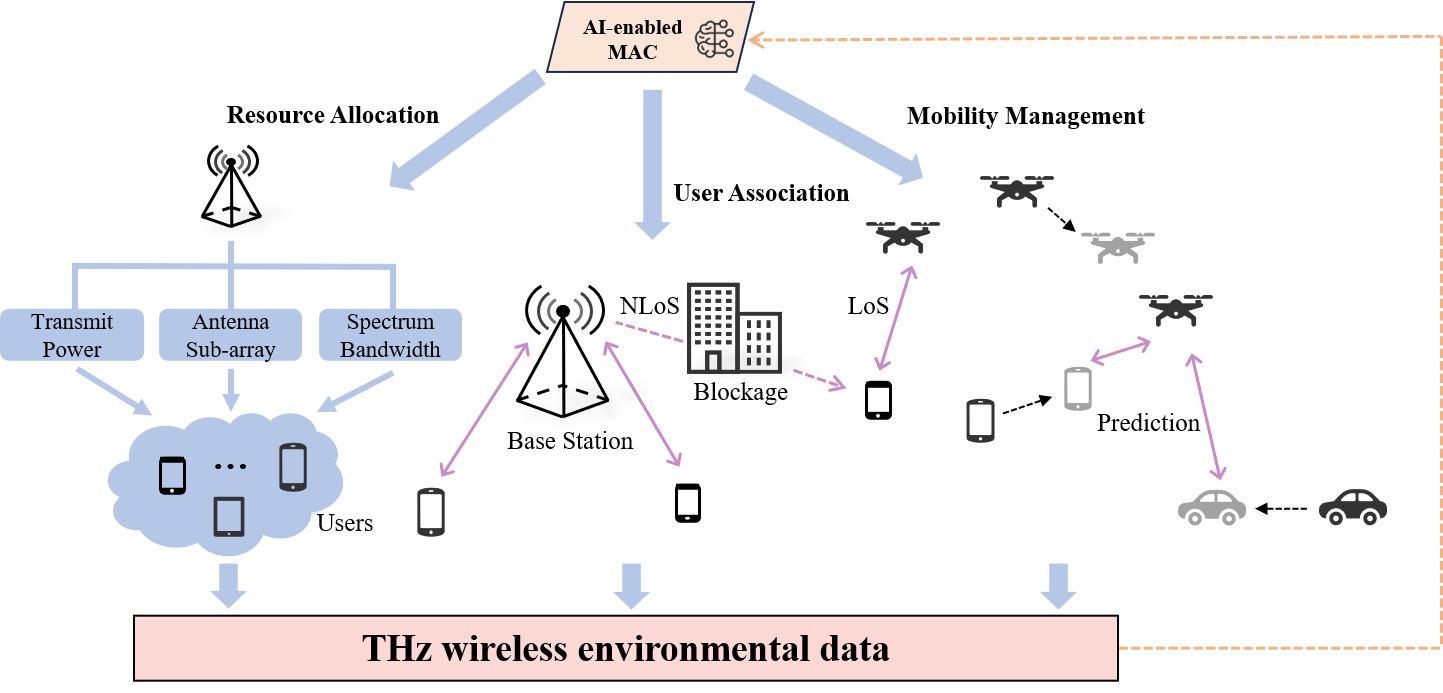}
    \caption{AI-enabled MAC layer management framework.}
    \label{fig_MAC}
    \vspace{-3mm}
\end{figure}

\vspace{-3mm}
\subsection{MAC Layer}
The MAC layer interfaces the physical and upper layers while managing user access. This process optimizes user association, resource allocation, and mobility to enhance network throughput, fairness, and energy efficiency. However, THz communications introduce critical challenges for MAC design. First, the multi-dimensional resource landscape comprising ultra-broad bandwidths and massive antenna arrays creates formidable complexity for traditional allocation algorithms. Second, highly directional beams and severe propagation loss make links vulnerable to dynamic environments and user mobility. Consequently, static management paradigms fail to satisfy dynamic resource demands. Third, the exponential growth of connected devices severely strains the scalability of conventional protocols \cite{10537167}. Furthermore, emerging deployment scenarios including SAGIN \cite{10679787}, digital twins, and multi-band integration increase network heterogeneity. Traditional fixed-rule MAC protocols lack the adaptability to such environments. Satisfying the distinct QoS requirements across URLLC, eMBB, and mMTC scenarios remains a substantial challenge for conventional user access schemes \cite{cui2025overview}.

To address these challenges, AI techniques, particularly DL and RL, enable the intelligent design of the MAC layer \cite{abusubaih2022intelligent, 8951180, 10368007}. As illustrated in Fig.~\ref{fig_MAC}, AI models can learn user mobility and environmental dynamics to predict viable access opportunities and candidate beams, thereby supporting optimal user association and faster beam discovery in THz networks. Then, learning-based policies can enable dynamic multi-dimensional resource allocation to adapt to specific service requirements.
In the following, we detail these AI applications across various MAC functions, while Table \ref{tab_mac} summarizes the existing solutions for THz MAC layer design.

\begin{table*}[t!]
\centering
\caption{AI-enabled MAC designs for THz networks.}
\label{tab_mac}
\scriptsize
\setlength{\tabcolsep}{1.6pt}
\renewcommand{\arraystretch}{1.02}
\begin{tabularx}{\textwidth}{
|>{\centering\arraybackslash}m{0.8cm}
|>{\centering\arraybackslash}m{4cm}
|>{\centering\arraybackslash}m{1.45cm}
|>{\centering\arraybackslash}m{2.5cm}
|>{\centering\arraybackslash}X
|>{\centering\arraybackslash}m{1.9cm}|}
\hline
\textbf{Ref.} & \textbf{MAC Function} & \textbf{AI Task} &
\textbf{AI Method} & \textbf{Main Focus} &
\textbf{Metric} \\ \hline
\cite{karthik2024mlbeam}
& User association
& Estimation
& Random Forest
& Rank beam candidates from reference-signal received power
& Beam quality\\ \hline
\cite{ismath2021fastia}
& User association
& Optimization
& Deep contextual bandit
& Infer per-AP access beams from one user uplink pilot
& Beam quality \\ \hline
\cite{ismath2021neighbor}
& User association
& Optimization
& Deep contextual bandit
& Predict waking AP beams from neighbors' positions and beams
& Beam quality \\ \hline
\cite{boulogeorgos2022artificial}
& User association$\&$Resource allocation$\&$Mobility
& Classification
& PPO$\&$LSTM
& Blockage-aware user association, beam selection and handover
& Beam quality \\ \hline
\cite{zhang2023gnn}
& User association$\&$Resource allocation
& Optimization
& GNN
& Graph edges guide user association and power allocation
& Throughput \\ \hline
\cite{van2025network}
& User association
& Optimization
& MADDQN
& THz link model informs service-aware access utility
& Total utility \\ \hline
\cite{shafie2024spectrum}
& Resource allocation
& Optimization
& DNN
& Power and spectrum allocation across absorption windows
& Throughput \\ \hline
\cite{hu2024multi}
& Resource allocation
& Optimization
& DDPG
& Sub-bands, sub-arrays, and power allocation
& Throughput \\ \hline
\cite{le2024resource}
& Resource allocation
& Optimization
& DDPG
& Tune power, bandwidth, beamwidth, altitude for LoS links
& Throughput \\ \hline
\cite{he2022intelligent}
& User association$\&$Resource allocation$\&$Mobility
& Prediction
& GNN$\&$LSTM$\&$DRL
& Position prediction for neighbor discovery and beam scheduling
& Throughput \\ \hline
\cite{termehchi2024distributed}
& Resource allocation$\&$Mobility
& Optimization
& MAPPO
& UAV trajectory optimization and channel allocation
& Energy efficiency \\ \hline
\end{tabularx}
\vspace{-3mm}
\end{table*}

AI-assisted methods can first reduce the beam discovery overhead associated with large directional codebooks. 
In \cite{karthik2024mlbeam}, partial reference-signal power measurements are processed by random-forest models to rank beam candidates, after which only the top candidates are re-measured for final selection. Another approach in \cite{ismath2021fastia} reuses information already available during coordinated access. Specifically, a deep contextual bandit maps a single UE uplink channel estimate to parallel access-beam decisions at multiple access points (APs), which avoids repeated downlink sweeping. Similarly, the work in \cite{ismath2021neighbor} uses the locations and beam decisions of active APs as contextual information to predict the beam of a newly activated AP, reducing redundant sounding and access delay. 
One step further, AI solutions support maintaining reliable THz connectivity under mobility and blockage.
In \cite{boulogeorgos2022artificial}, user and blocker geometry is first exploited for user association and blockage avoidance, while temporal position and SINR information guide PPO-based beam selection. Supervised channel prediction then supports proactive handover before link degradation, improving THz access continuity.
Moreover, the framework in \cite{he2022intelligent} utilizes GNN and LSTM to predict network states from node mobility and link conditions. The predicted states identify viable neighbors, while nested DRL is responsible for codebook selection and beam adjustment to maintain narrow-beam alignment under rapid network dynamics.

After link establishment, AI can further facilitate efficient multi-dimensional resource allocation in THz networks. In \cite{hu2024multi}, the authors develop a DDPG-based algorithm with hybrid discrete and continuous actions to jointly allocate multiple THz resources, where sub-array and sub-band assignments are modeled as discrete variables and transmit power as a continuous variable. This joint design improves network throughput while maintaining user fairness. Similarly, the work in \cite{le2024resource} extends resource allocation to link geometry by treating HAP altitude and antenna beamwidth as controllable variables, so that directional gain and coverage can be coordinated with power and bandwidth in THz aerial networks.
However, the preceding schemes do not explicitly adapt sub-band allocation to the nonlinear frequency response induced by molecular absorption. The unsupervised schemes in \cite{shafie2024spectrum} 
incorporate absorption response into the training objective and learn unequal sub-bandwidth and power allocation accordingly.

In response to the increasing network heterogeneity and diverse service requirements, AI-enabled MAC protocols are tailored for complex scenarios. 
For instance, the work in \cite{termehchi2024distributed} proposes a distributed MADRL algorithm for UAV-assisted aerial networks. This solution controls UAV trajectories according to user locations and THz rate satisfaction, while channel assignment is performed based on the resulting link rates.
In the digital-twin network of \cite{zhang2023gnn}, THz channel gains determined by spreading and molecular absorption form graph features for joint user association and power allocation, allowing the GNN to scale its decisions with network topology.
Furthermore, to accommodate multi-band heterogeneous environments, a MADRL framework in \cite{van2025network} facilitates decentralized network selection across sub-6~GHz, mmWave, and THz bands. This enables diverse eMBB and URLLC users to satisfy distinct service requirements.

\vspace{-3mm}
\subsection{Network Layer}
Routing is a fundamental network layer function that determines paths from source to destination to ensure efficient packet delivery. Traditional algorithms utilizing shortest-path or distance-vector metrics, including OSPF \cite{moy1997ospf} and RIP \cite{malkin1998rip}, suffer from ineffective decisions and degraded resource utilization in large-scale heterogeneous environments \cite{yang2022comparative, tera_ref}. THz networks further introduce unique routing challenges. First, highly dynamic topologies cause frequent link failures. This requires low-complexity algorithms for rapid path recovery. Second, the higher path loss of THz links necessitates multi-hop relaying to extend coverage, thereby complicating routing design. Third, ultra-high data rates severely strain the memory buffers of network devices. Lastly, diverse deployment scenarios prioritize distinct performance metrics. For example, resource-constrained nano-networks demand strict energy efficiency. Traditional static metrics fail to support this essential multi-objective optimization.


Instead, AI techniques enable adaptive and efficient routing for THz networks. Especially, Q-learning methods can incorporate directional link availability, buffer occupancy, and other resource constraints into their states or rewards, while dynamically selecting feasible next hops according to current network conditions. In this framework, each node evaluates candidate relays using such THz-aware feedback and can retain alternative forwarding actions when the preferred link becomes temporarily unavailable. This enables the routing protocol to adapt to unexpected THz network failures, such as buffer overflow, thereby maintaining reliable routing performance.
We review recent DRL-based routing designs as follows and summarize the findings in Table \ref{tab_network}.

In \cite{xia2020routing}, Q-learning adapts next-hop selection to intermittent directional THz connectivity. Antenna facing time and residual buffer capacity are incorporated into the reward to avoid selecting short paths with temporarily unavailable directional links, achieving high packet arrival rates and minimal buffer blockage. 
To address greater dynamics and tighter latency constraints in SAGIN, the study in \cite{hassan2024spaceris} employs multi-agent PPO for LEO satellite routing, while a separate learning agent adapts RIS phases to the cascaded THz channel to enhance link quality along the selected paths. 
Furthermore, the Q-learning scheme in \cite{deb2021xia} addresses range-induced disconnections in THz UAV networks. Packets are forwarded through UAV relays when feasible multi-hop THz paths exist and switched to LEO satellites otherwise, with THz transmission delay guiding routing.

\begin{table*}[t!]
\centering
\caption{AI-based routing designs for THz networks.}
\label{tab_network}
\scriptsize
\setlength{\tabcolsep}{1.7pt}
\renewcommand{\arraystretch}{1.03}
\begin{tabularx}{\textwidth}{
|>{\centering\arraybackslash}m{0.9cm}
|>{\centering\arraybackslash}m{1.0cm}
|>{\centering\arraybackslash}m{1.45cm}
|>{\centering\arraybackslash}m{1.3cm}
|>{\centering\arraybackslash}X
|>{\centering\arraybackslash}m{5.00cm}|}
\hline
\textbf{Ref.} & \textbf{Scenario} & \textbf{AI Task} &
\textbf{AI Method} & \textbf{THz-Specific Constraint / Decision} &
\textbf{Routing Metric} \\ \hline
\cite{xia2020routing}
& \shortstack{General}
& Optimization
& Q-learning
& Select/deflect next hops considering narrow beams and buffers
& Beam directions$\&$Buffer occupancy$\&$Link traffic \\ \hline
\cite{hassan2024spaceris}
& \multirow{2}{*}{\shortstack{SAGIN}}
& Optimization
& MAPPO
& Select LEO routes and RIS phases for dynamic sub-THz links
& Throughput$\&$Packet-delivery delay \\ \cline{1-1}\cline{3-6}
\cite{deb2021xia}
&
& Optimization
& Q-learning
& Select UAV relays or LEO fallback when short-range THz paths disconnect
& Packet-delivery delay \\ \hline
\cite{hu2023deep}
& \shortstack{Backhaul}
& Optimization
& MADDPG
& Select distance-aware paths, then allocate power and Tx/Rx sub-arrays
& Recovery$\&$Resource efficiency \\ \hline
\cite{alshorbaji2024energy}
& \multirow{3}{*}{\shortstack{Nano-\\network}}
& Optimization
& Q-learning
& Select relay/bandwidth under short range and nanocapacitor limits
& Energy consumption \\ \cline{1-1}\cline{3-6}
\cite{garcia2023dynamic}
&
& Optimization
& Q-learning
& Drop, store, or transmit as blood flow changes THz contact opportunities
& Throughput$\&$Coverage$\&$Energy consumption \\ \cline{1-1}\cline{3-6}
\cite{wang2020multi}
&
& Optimization
& Online RL
& Deflect to backup relays using harvested energy and ACK feedback
& Buffer occupancy$\&$Energy consumption \\ \hline
\end{tabularx}
\vspace{-5mm}
\end{table*}

Beyond these works, multi-hop routing is essential to extend THz network coverage. For mesh backhaul networks, the authors in \cite{hu2023deep} develop a cross-layer design that first selects routes using a THz distance-dependent resource cost and then employs hierarchical multi-agent DDPG to allocate power and antenna sub-arrays according to link SINR and buffer occupancy. This strategy facilitates rapid policy adaptation after link failures without reconstructing the learning architecture.
Moreover, applying multi-hop routing to nanoscale networks introduces severe resource and buffer constraints. 
In \cite{alshorbaji2024energy}, Q-learning actions jointly select the next relay and transmission bandwidth, so that the routing policy can balance transmission energy against the additional hops required by short-range THz links. This problem becomes more dynamic when nanodevices move with the surrounding medium. 
In \cite{garcia2023dynamic}, the node accordingly decides whether an event should be transmitted immediately or temporarily stored until a feasible forwarding opportunity emerges. When the preferred relay is unavailable because of energy depletion, buffer occupancy, or link errors, \cite{wang2020multi} further introduces deflection routing to redirect packets through alternative nanonodes.

\vspace{-3mm}
\subsection{Transport Layer}
The transport layer ensures reliable end-to-end communication and efficient congestion control. Nevertheless, conventional protocols like Transmission Control Protocol (TCP) struggle in THz networks. First, although THz links support extreme data rates under ideal conditions \cite{polese2020toward}, they introduce tremendous traffic fluctuations. Traditional TCP \cite{allman2009tcp} suffers from slow start and utilizes fixed rules to adjust the congestion window. This rigid mechanism fails to adapt to rapid traffic variations and ultimately causes congestion. Regarding reliable transmission, the inherent vulnerability of THz links causes frequent outages. The transport layer must subsequently execute repeated retransmissions, which increases packet overhead and exacerbates congestion. Furthermore, these outages trigger retransmission timeouts that drastically reduce throughput \cite{khorov2023boosting}. Finally, conventional TCP primarily relies on packet loss \cite{mo1999analysis} and round-trip time \cite{brakmo1994tcp} to perceive network conditions. These metrics may not reflect actual congestion states \cite{tang2021survey}. As THz networks scale and become more heterogeneous, accurately assessing network conditions becomes a formidable challenge for traditional protocols.

To overcome these limitations, AI-driven methodologies are being explored to enhance transport layer intelligence and adaptability. For instance, AI technologies enable precise traffic prediction to anticipate network dynamics. Specifically, the authors in \cite{qiu2018spatio} combine LSTM and multi-task learning to analyze traffic characteristics across individual and neighboring cells for accurate forecasting. 
One step further, considering the impact of external factors, an approach combining DL and transfer learning in \cite{zhang2019deep} extracts spatial-temporal features and cross-domain data, including base station distributions and social activities, to achieve high-precision traffic prediction. 
On the other hand, DRL approaches enable intelligent congestion control. The authors in \cite{li2016learning} utilize Q-learning to adaptively control the congestion window by evaluating ACK intervals, packet transmission intervals, and RTT ratios. Similarly, an asynchronous advantage actor-critic algorithm (A3C) in \cite{nie2019dynamic} optimizes the initial congestion window and executes real-time control decisions based on user-specific network features. 
Despite these advancements, research dedicated to the THz transport layer remains in its infancy. 
Unique THz network characteristics exacerbate traffic prediction difficulties and impose stringent demands on reliability, while existing AI-based traffic prediction and congestion control schemes may not readily adapt to THz transport dynamics. Building on these existing approaches, future THz transport protocols can incorporate cross-layer information, such as beam and link states, together with short-term capacity prediction to enable faster adaptation of congestion control to rapid link variations.

\vspace{-3mm}
\subsection{Performance Metrics and Lessons Learned}
For higher-layer THz networking, we summarize several network performance metrics that can serve as references for evaluating and optimizing AI-assisted protocols. 
At the MAC layer, user association and resource allocation are commonly evaluated by throughput or data rate, spectral and energy efficiency, fairness, and QoS satisfaction. 
As the focus moves to the network layer, packet loss and end-to-end delay become dominant, together with buffer blockage, hop count, resource efficiency, and recovery time after link failures. 
At the transport layer, end-to-end delivery and congestion control are mainly characterized by RTT, queuing delay, retransmission timeout, and fairness. Additionally, several lessons can be drawn from the reviewed studies.
\begin{itemize}
\item \textbf{Cross-layer design.}
THz introduces distinct challenges across different protocol layers. Specifically, the MAC layer faces multi-dimensional resource allocation and blockage-sensitive user access, the network layer must accommodate limited buffer capacity and rapid path recovery, and the transport layer must adapt to highly dynamic traffic conditions. However, THz characteristics, like directional beams, simultaneously influence multiple protocol layers, motivating cross-layer AI to jointly coordinate decisions and manage the networks.

\item \textbf{AI-native multi-objective optimization.}
Current solutions largely treat functions such as resource allocation, mobility prediction, and traffic forecasting as separate modules, although their objectives and decisions are inherently coupled. Future AI-native networking should integrate these functions into unified multi-objective learning and control frameworks to enable coordinated network-wide optimization.
\end{itemize}

\section{AI for THz Service: Computing and Sensing}
\label{sec_service}
Transformative applications such as the Metaverse and XR demand unprecedented data throughput and ultra-low latency. To meet these requirements, mobile edge computing (MEC) migrates complex processing tasks to the network edge to minimize service latency. Concurrently, THz sensing captures precise environmental information through high-precision localization and reconstruction. Serving as the fundamental technology, THz communication ensures the transmission of the massive data streams generated by these services. 
The deep integration of communication, computing, and sensing provides the cornerstone for these emerging network services. 
Furthermore, AI techniques efficiently coordinate this integration to elevate overall service performance and user experiences.
Consequently, this section presents a detailed survey of AI technologies applied to THz MEC, THz sensing, and integrated sensing, computing and communication (ISCC).

\begin{table*}[t!]
\centering
\caption{Summary of AI Methods in THz Sensing}
\label{tab_thz_sensing}
\scriptsize
\setlength{\tabcolsep}{1.7pt}
\renewcommand{\arraystretch}{1.03}
\begin{tabularx}{\textwidth}{|>{\centering\arraybackslash}m{1.5cm}|>{\centering\arraybackslash}m{2.8cm}|>{\centering\arraybackslash}m{1.4cm}|>{\centering\arraybackslash}m{3cm}|>{\centering\arraybackslash}X|>{\centering\arraybackslash}m{2cm}|>{\centering\arraybackslash}m{0.7cm}|}
\hline
\textbf{Sensing Task} & \textbf{Feature Input} & \textbf{AI Task} & \textbf{AI Method} & \textbf{Main Focus} & \textbf{Metric} & \textbf{Ref.} \\ \hline
\multirow{5}{*}{Localization}
& CSI 
& Estimation
& LSTM$\&$Attention
& Multipath-resilient 3D positioning
& Localization error
& \textbf{\cite{fan2021siabr}} \\ \cline{2-7}
& CSI
& Estimation
& CNN$\&$Attention
& Blockage-robust indoor positioning
& Localization error
& \textbf{\cite{chiu20246g}} \\ \cline{2-7}
& Sub-array signals
& Estimation
& Transformer
& Parallax-aided near-field positioning
& Localization error
& \textbf{\cite{zeng2026Parallax}} \\ \cline{2-7}
& Event/time reports
& Classification
& GNN
& Flow-guided event localization
& Region accuracy
& \textbf{\cite{bartra2024graph}} \\ \cline{2-7}
& Angle/delay estimates
& Prediction
& RNN$\&$Transformer
& Blockage-resilient user tracking
& Imputation error
& \textbf{\cite{chaccour2024joint}} \\ \hline
\multirow{3}{*}{\shortstack{Environment\\Reconstruction}}
& Scattering points
& Generative
& Transformer
& 3D object reconstruction
& Volume overlap
& \textbf{\cite{chen2024dlter}} \\ \cline{2-7}
& Received signal strength
& Generative 
& Conditional GAN
& Obstacle recovery from blockage shadows
& Average precision
& \textbf{\cite{hu2024towards}} \\ \cline{2-7}
& Received signal strength
& Generative
& Conditional GAN$\&$Voting
& Directional map and obstacle recovery
& Average precision
& \textbf{\cite{hu2025advancing}} \\ \hline
\end{tabularx}
\vspace{-3mm}
\end{table*}

\vspace{-2mm}
\subsection{THz Edge Computing}
The proliferation of smart devices and complex services demands immense computational and transmission capabilities. MEC alleviates device-side computation by offloading tasks to nearby edge servers, while the THz links can substantially reduce task transmission latency and make edge execution feasible for data-intensive applications. In general, THz MEC requires task offloading to be coordinated with communication and computing resources according to both network dynamics and server availability. AI methods, particularly DRL, provide a practical approach to learning such coupled decisions under time-varying conditions. For instance, the work in \cite{du2020mec} employs A3C to jointly determine rendering task offloading and downlink power allocation based on dynamic user requests, thereby reducing long-term energy consumption.

However, the performance of the MEC network is highly dependent on THz link conditions, which are sensitive to the environment. Considering the mobility scenarios, the work in \cite{zhao2023mobility,zhao2024mobility} first performs distance-aware THz bandwidth allocation to account for spectrum availability. Then offline RL schemes are proposed to optimize task-offloading decisions to reduce latency and energy consumption while avoiding unsafe online trial-and-error interactions. To actively improve THz connectivity, UAV-assisted MEC exploits aerial mobility to shorten transmission distances and maintain LoS links. The studies in \cite{li2024energy,wang2022joint,ren2023joint,park2023joint} accordingly coordinate UAV placement or trajectories with task offloading and resource allocation through DRL algorithms, allowing computing decisions to adapt to the resulting variations in THz link rate and offloading latency. Another direction is to mitigate weak or blocked links through RIS. The frameworks in \cite{wu2023fast,wu2024two} adopt imitation-based offline policy optimization frameworks to learn task offloading decisions and optimize RIS phase configurations to improve the corresponding THz links. Furthermore, cell-free networking provides a distributed architecture, which also alleviates the short coverage and blockage sensitivity of THz access. A MAPPO algorithm in \cite{jiang2026multi} jointly optimizes task offloading and multi-dimensional THz communication resources to balance the latency-energy tradeoff.

Beyond energy and latency, other metrics are critical for specific applications. 
For virtual reality applications, the constrained DRL framework in \cite{liu2021learning} predicts user mobility and LoS conditions for RIS configuration, thereby improving downlink rates and user QoE. 
Moreover, in resource-constrained SAGIN networks, the authors in \cite{hu2024tera} propose a GNN-based DRL approach to optimize task offloading between satellites and enhance resource utilization under time-varying network topology. In \cite{mao2021optimizing}, the authors employ an LSTM to predict harvested energy and the predictions guide offloading decisions to maximize task success rates, while the THz link primarily provides short-range access.

\vspace{-2mm}
\subsection{THz Sensing}
THz ISAC has emerged as a key paradigm for future wireless networks. Benefiting from the ultra-broad bandwidth and sub-millimeter wavelength of the THz band, this integration enables high-resolution sensing capabilities concurrent with data transmission. 
By processing received wireless signals, THz sensing extracts target physical parameters, such as range, velocity, and angle. This capability supports diverse services, enabling high-precision localization, environment reconstruction, moving target detection, etc.
While traditional signal processing methods provide fundamental solutions for these tasks, AI techniques demonstrate significant potential to enhance the accuracy of sensing tasks through their powerful data feature extraction \cite{luong2025advanced} and to optimize QoE in specific THz network applications, particularly XR \cite{chaccour2024joint}.
Subsequently, we provide a detailed review of the application of AI methods in critical THz sensing domains such as localization and environment reconstruction, which are summarized in Table~\ref{tab_thz_sensing}.

THz localization faces critical challenges from severe path loss, diffuse scattering, and channel sparsity. Traditional geometric ranging fails in cluttered indoor environments with non-line-of-sight conditions and severe multipath interference. Conversely, DL effectively extracts implicit features from CSI, and attention mechanisms can further mitigate these interference impacts. 
In \cite{fan2021siabr}, an attention-based bidirectional LSTM identifies informative components from sparse and irregular THz multipath observations, thereby suppressing unreliable propagation paths for precise indoor localization. Similarly, the work in \cite{chiu20246g} employs CNNs with attention mechanisms to transform localization into pattern matching based on CSI to significantly enhance three-dimensional accuracy.
In near-field scenarios, the transformer-based approach in \cite{zeng2026Parallax} fuses spatially diverse observations across sub-arrays to facilitate efficient positioning.
Furthermore, THz localization extends to health monitoring applications. The study in \cite{bartra2024graph} applies GNNs to localize intravascular nanodevices by capturing complex anchor-region relationships within vascular maps, thereby reducing positioning errors under blood flow guidance.

In addition to localization, THz ultra-wide bandwidth enables high-resolution environment reconstruction from echo signals. However, channel sparsity generates uneven sensing data, hindering global environmental characterization. Traditional simultaneous localization and mapping techniques yield only two-dimensional or sparse point clouds, struggling to precisely reconstruct complex three-dimensional urban environments. 
To address these challenges, advanced DL models like transformers effectively process sparse THz scattering data to capture complex distributions. 
Specifically, the authors in \cite{chen2024dlter} employ a point transformer to capture spatial relationships from incomplete and irregular scatterer point clouds caused by THz scattering. By decomposing the reconstruction process into object detection and shape generation, the framework can recover accurate three-dimensional object models from these sparse observations. Alternatively, GANs in \cite{hu2024towards} learn THz attenuation under blockage to infer complete signal maps from sparse data and extract obstacle positions by identifying shadow edges. Furthermore, the study in \cite{hu2025advancing} exploits the direction-dependent blockage patterns induced by narrow THz beams and diffraction. By jointly learning directional radio maps and aggregating obstacle evidence across different beam directions, the framework improves both radio map construction and obstacle sensing.

\vspace{-3mm}
\subsection{Integrated Sensing, Computing and Communication}
\begin{figure}[t!]
    \centering
    \includegraphics[width=0.95\linewidth]{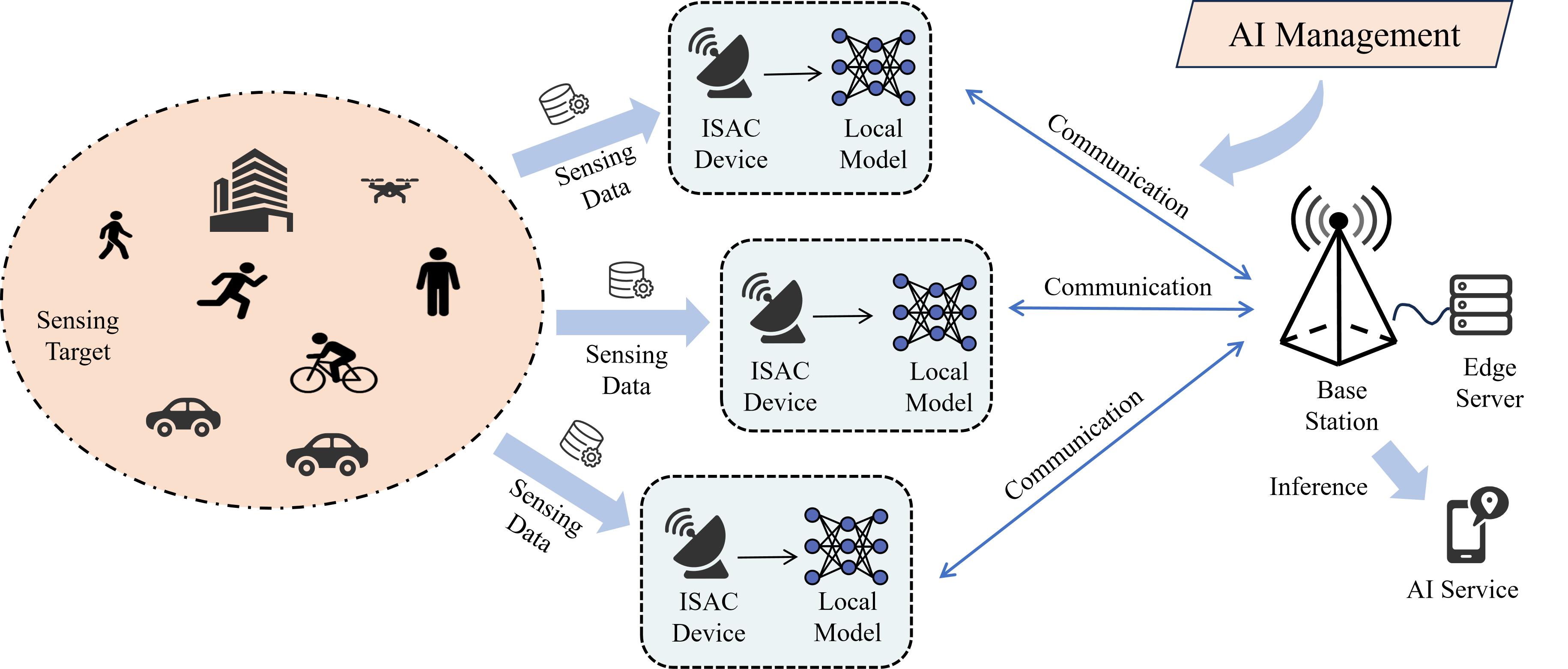}
    \caption{The system architecture of an ISCC framework.}
    \label{fig_iscc}
    \vspace{-3mm}
\end{figure}

6G networks aim to foster an intelligent society by integrating sensing, communication, computation, and AI into a unified ISCC framework \cite{wu2025ai}, as illustrated in Fig.~\ref{fig_iscc}. Within this paradigm, the THz band provides a physical foundation by offering ultra-high throughput and millimeter-level sensing resolution. Concurrently, AI dynamically orchestrates heterogeneous network resources. In turn, by acquiring high-fidelity sensing data, offloading it via high-speed THz links, and leveraging edge computing, ISCC constructs a formidable infrastructure for large-scale AI training and inference.

AI techniques have been extensively deployed to optimize ISCC system performance. For instance, a DDPG algorithm in \cite{yang2024deep} jointly optimizes beamforming and power allocation in highly dynamic vehicular networks. This approach maximizes the achievable data rate while strictly guaranteeing sensing and computational performance. 
Similarly, the work in \cite{feng2022joint} addresses the issues of high latency and low resource utilization prevalent in machine-type communication networks. By leveraging DL and RL paradigms to process massive sensing data, the proposed framework intelligently optimizes resource allocation, routing, and waveforms to facilitate ultra-low-latency closed-loop control.

The ISCC paradigm also facilitates AI model training and inference. Given distributed network data and privacy constraints, FL has emerged as a solution. Within integrated systems, some research efforts have focused on accelerating model convergence and enhancing inference accuracy. 
The authors in \cite{wen2025integrated} investigate an over-the-air FL scenario where edge devices locally train models using wireless sensing data and aggregate local gradients via over-the-air computation. This work alternately optimizes training batch sizes and network resource allocation to maximize convergence speed under strict latency and energy budgets. 
In a similar task-oriented context, a closed-loop adaptive strategy in \cite{liu2022toward} trains local neural networks using radar spectrogram data. By dynamically adjusting batch sizes based on loss evolution and jointly allocating resources, this system significantly accelerates federated convergence.
Furthermore, to elevate task accuracy under integrated operations, the study in \cite{du2024integrated} mitigates interference between gradient aggregation signals and radar echoes. By jointly optimizing transceiver beamforming and device selection, this approach minimizes transmission errors while satisfying sensing demands. For edge-device collaborative inference, the authors in \cite{zhuang2023integrated} split the AI model across the network. Devices extract low-dimensional features from sensing signals and transmit them via over-the-air computation to maximize discrimination gain, which effectively suppresses noise under limited communication resources.

\vspace{-2mm}
\subsection{Performance Metrics and Lessons Learned}
For THz-enabled services, performance metrics depend closely on the computing and sensing task. 
In THz MEC networks, energy consumption and service latency are the primary metrics, while QoE and computing task success rate are further considered for application-specific objectives. 
For THz sensing, localization performance is commonly evaluated by position or angle estimation errors, and environment reconstruction is typically characterized by reconstruction MSE and target detection precision.
Several observations follow.

\begin{itemize}
\item \textbf{Physics-aware sensing}. 
THz sensing provides observations such as sparse high-resolution multipath, near-field parallax, irregular scatterer point clouds, and blockage-shaped radio maps. 
Embedding these structures into AI representations and models is preferable to directly applying generic learning architectures to THz sensing data.

\item \textbf{Bidirectional relationship between AI and ISCC}. 
AI jointly optimizes sensing, communication, and computing resources, while ISCC provides the distributed sensing and computing pipeline required for intelligent inference and learning. Extending to THz networks, the large link capacity can support more frequent model aggregation and richer sensing-feature exchange, thereby accelerating convergence and enabling more flexible model partitioning across devices and edge servers. However, how to realize environment-aware collaboration and adaptive model partitioning remains an open problem.
\end{itemize}

\section{THz Network for AI Service}
\label{sec_thz4ai}
Benefiting from the continuous evolution of wireless networks that integrate communication, computation, and sensing, the network infrastructure is expected to support the deployment, training, and inference of AI models. This transformative paradigm is widely recognized as the Network for AI \cite{cui2025overview}.
The preceding sections demonstrated AI empowering THz systems, yet their relationship is fundamentally symbiotic. 
Beyond providing ultra-wide bandwidth and low transmission latency for communication-intensive AI training and inference, reconfigurable THz links can adapt network connectivity to distributed AI workloads and computation demands. 
Furthermore, high-resolution THz sensing can acquire high-quality data from the physical world to support AI model training, notably empowering embodied AI through enhanced environmental perception and interaction.
At a deeper level, emerging THz physical computing architectures allow EM propagation to participate in AI processing, extending the role of THz from information transmission and acquisition to computation.
Driven by this reciprocal relationship, this section reviews THz networks for AI training, inference, data collection, and THz physical computing paradigms for AI.


\vspace{-2mm}
\subsection{THz Network for AI Training}
The unprecedented evolution of generative AI and LLMs has fundamentally transformed computational paradigms. Modern AI models encompass billions of parameters and require training across massive datasets, which exceeds the processing capabilities of any single computing node. 
Consequently, distributed machine learning has become indispensable. During distributed training, thousands of participating nodes must continuously synchronize massive gradients, activations, and weights through collective communication operations, such as AllReduce and All-to-All, while distributing segments of vast datasets.
This intensive synchronization generates enormous communication traffic, which hinders overall training efficiency. 
Existing solutions \cite{alistarh2017qsgd,lin2018deepgradient,vogels2019powersgd} mainly reduce network communication overhead through gradient compression.
However, aggressive compression may introduce gradient approximation errors, additional encoding and reconstruction costs, or convergence degradation. 
Instead, THz communications provide a transformative solution for distributed AI training by expanding the available communication capacity and reducing parameter synchronization latency.

Data centers provide the primary infrastructure for large-scale AI training, but conventional fiber or copper rack interconnects offer limited post-deployment scalability and flexibility for dynamic training traffic. Wireless data centers have therefore been investigated to provide flexible rack-to-rack connectivity \cite{hamza2016wireless}. Short-range THz links can offer higher capacity and nanosecond-level latency with higher energy efficiency than wired interconnects \cite{han2025wires}. By supplementing cables with on-demand links, THz connectivity also improves interconnect flexibility and reduces cabling overhead. Recent studies have examined the physical feasibility of this architecture. The work in \cite{eckhardt2024hybrid} investigates point-to-point THz rack-to-rack channel characteristics, while \cite{zhao2025terahertz} explores near-field beamforming to improve the spectral efficiency of wireless data centers. These studies provide a physical foundation, but realizing training acceleration requires further investigation of network scheduling tailored to distributed learning.

Existing studies on distributed training show that communication scheduling should account for computation dependencies and critical-path effects, since different tensor transfers contribute unequally to iteration completion time \cite{jayarajan2019p3,hashemi2019tictac,peng2019bytescheduler}. Reconfigurable THz links can extend this principle to the physical network by establishing temporary direct connections for transfers that currently limit training progress. Beyond transmission scheduling, network infrastructure can also participate in training operations. The work in \cite{sapio2021switchml} aggregates gradients within programmable switches to reduce repeated end-host communication. A similar principle could be implemented through THz over-the-air computation \cite{che2026photonics}, supporting gradient and parameter aggregation. Furthermore, collective communication performance depends on topology, routing, and transmission scheduling \cite{shah2023taccl,wang2023topoopt,liu2024teccl}. The hybrid optical-THz AI data center in \cite{jiang2026thzsync} builds on this coupling by complementing persistent optical interconnects with a reconfigurable THz overlay. Its framework synthesizes collectives based on their dependencies while jointly allocating THz resources and coordinating cross-fabric traffic in response to congestion, channel variations, and straggler states. This coordination reduces collective communication completion time and transmission energy consumption.

Rather than relying exclusively on data centers, AI training is progressively migrating toward the network edge. In edge networks, massive distributed terminals such as autonomous vehicles, UAVs, and smartphones participate as worker nodes in distributed training. While THz links provide the necessary capacity to minimize dataset and gradient transmission delays, deploying distributed training across the edge introduces formidable challenges. These difficulties primarily stem from the severe heterogeneity of user computational capabilities, highly dynamic wireless channels, and the non-independent and identically distributed nature of local datasets. Efficiently scheduling network resources to accommodate these diverse constraints and accelerate model convergence requires highly sophisticated management strategies.

To navigate these intricacies, recent literature has explored how wireless networks can effectively support distributed learning paradigms. 
A prominent direction is the hybrid distributed learning architecture, which dynamically assigns FL or split learning roles to users to optimally balance local computational burdens and communication overheads. 
The authors in \cite{liu2023distributed} developed a hybrid framework that evaluates heterogeneous user capabilities regarding compute power, dataset size, and channel quality. By jointly optimizing wireless resource allocation and selecting participants based on model update importance, this approach achieves superior learning performance. Similarly, the study in \cite{liu2022wireless} actively tailored individual UAV training roles according to their specific communication and computation limitations. 
Moreover, the work in \cite{tao2024federated}  elaborated on learning-task-oriented communication designs. Rather than relying on traditional metrics like throughput, this methodology establishes learning objectives, including convergence accuracy and speed, as the primary optimization criteria for wireless system design. This task-oriented approach effectively overcomes data and resource heterogeneity to ensure highly robust distributed learning.

\begin{figure}[t!]
    \centering
    \includegraphics[width=0.95\linewidth]{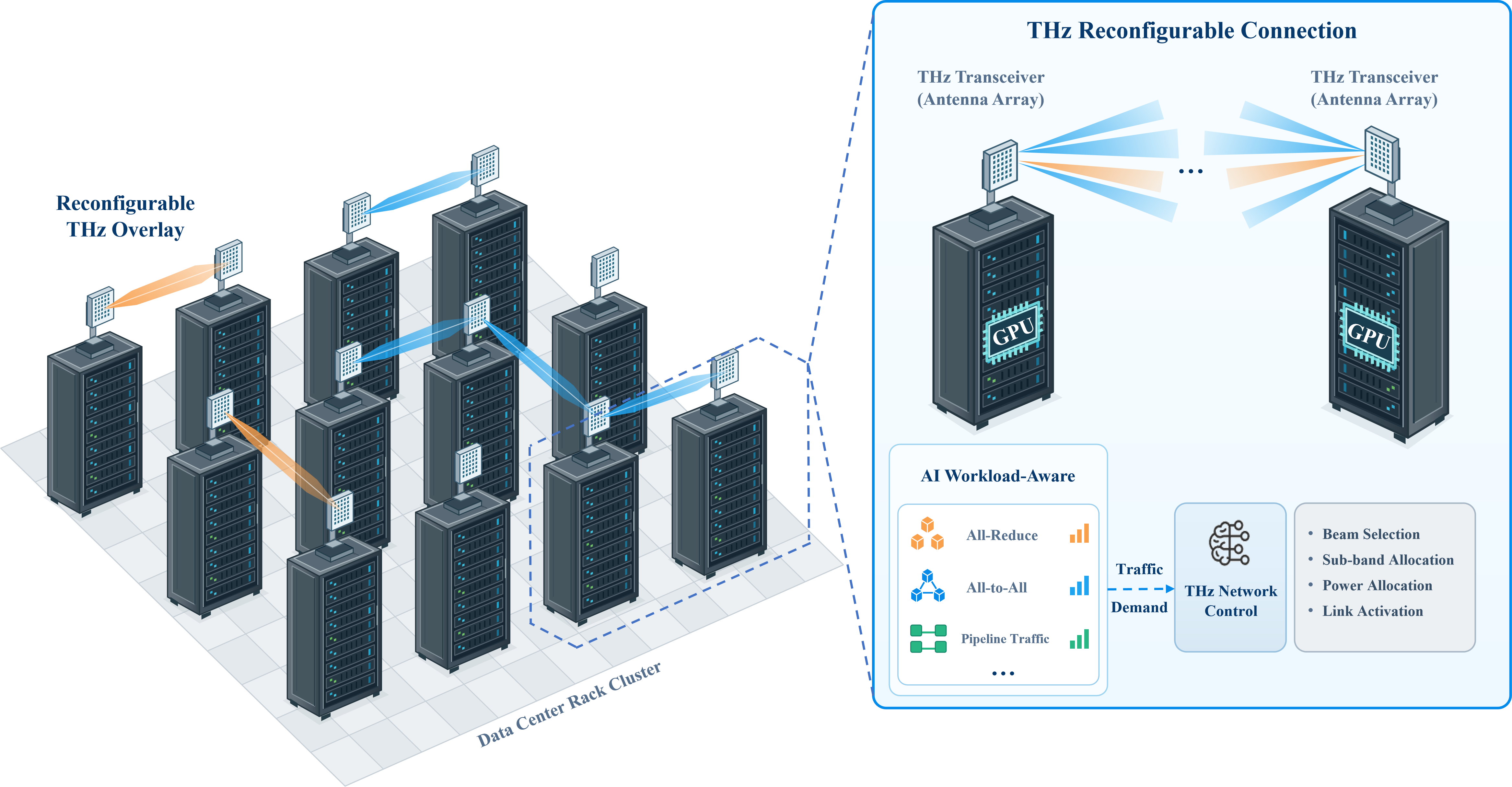}
    \caption{THz-Enabled Reconfigurable Wireless AI Data Center.}
    \label{fig_thz4ai1}
    \vspace{-3mm}
\end{figure}

\vspace{-2mm}
\subsection{THz Network for AI Inference}
The proliferation of highly demanding AI applications, such as autonomous driving and LLMs, has reshaped the landscape of modern technological services. 
These advanced AI applications inherently rely on sophisticated models characterized by massive parameter scales and formidable memory consumption. 
Therefore, deploying and executing these intensive inference tasks directly on resource-constrained terminal devices presents profound difficulties. 
Furthermore, these applications impose extremely stringent latency constraints. 
THz communications provide the essential transmission infrastructure to support rapid inference task offloading and inter-node coordination, effectively resolving the severe high latency and communication congestion issues.

Recent wireless networking research establishes foundational methodologies for efficient AI inference that could be extended to THz networks. The resource allocation scheme in \cite{ahmad2026resource} assigns inference tasks to nearby computing nodes in multi-user wireless sensing edge networks to reduce latency and energy consumption. The study in \cite{he2025large} further considers heterogeneous edge and cloud resources and jointly determines server selection, computational and communication resource allocation for efficient inference execution. These studies primarily consider complete task offloading, whereas model partitioning retains part of the computation at the terminal. In \cite{li2019edge}, DNN execution is split between the device and edge server, which receives intermediate features from locally executed front-end layers for subsequent processing. To adapt to computation and communication conditions, \cite{kang2017neurosurgeon} dynamically selects the partition point based on layer execution costs, intermediate feature sizes, network conditions, and server load. The framework in \cite{eshratifar2021jointdnn} extends this approach by mapping multiple model segments between the device and cloud under computation, communication, energy, and latency constraints.

Another critical paradigm for executing complex AI models over wireless networks is collaborative inference. 
The framework in \cite{shlezinger2023collaborative} supports hierarchical partitioning of large DNNs or prediction aggregation from independently executed compact models. In multi-user wireless networks, \cite{li2024optimal} enables devices to execute front-end layers locally and concurrently upload intermediate features to an edge server, which completes inference and combines results. Finer-grained partitioning can further increase parallelism. The work in \cite{zhao2018deepthings} divides convolutional computation into spatial tiles dynamically assigned to resource-constrained edge devices, while \cite{hu2022edgeflow} decomposes general DNN computation graphs into fine-grained execution units. Thus, inference granularity and placement depend jointly on computing and network heterogeneity.

The emergence of generative AI introduces additional communication requirements for distributed inference, particularly for LLM serving in data centers. 
Unlike conventional DNN inference, autoregressive LLMs maintain persistent intermediate states and exhibit distinct resource demands during the prefill and decoding stages. 
The work in \cite{zhong2024distserve} separates these phases across different GPU groups and selects parallelization and placement based on computation requirements and inter-node bandwidth, with the key-value (KV) cache transferred between stages. 
Similarly, the study in \cite{patel2024splitwise} assigns the two inference phases to heterogeneous resource pools and transfers KV states progressively across model layers so that communication can overlap with ongoing computation. 
Going beyond static placement, the authors in \cite{sun2024llumnix} migrate active inference requests and their KV states across serving instances during execution to respond to load imbalance and memory pressure while limiting inference interruption. 
In such scenarios, the reconfigurable THz links can accelerate KV-cache transfer and dynamically connect distributed accelerator pools, thereby reducing inter-stage communication latency and improving LLM serving flexibility and efficiency.

Finally, efficient inference also requires robustness to wireless connectivity variations.
The work in \cite{shao2020bottlenet} inserts trainable feature encoders and decoders around a wireless channel and optimizes the transmitted representation directly for the final inference objective. This approach enables transmission to adapt to different channel conditions.
Another study in \cite{laskaridis2020spinn} jointly adapts model partitioning and early inference exits according to network and computing conditions. 
When remote execution becomes unavailable, a local exit can still provide an inference result rather than causing the entire task to fail. 

\subsection{THz Network for Data Collection}
The effectiveness and generalization capabilities of advanced AI models fundamentally rely on the scale and quality of their training data. 
THz networks are capable of acting as real-time data collectors. 
From a networking perspective, the continuous operation of a THz system produces a wealth of implicit data. 
By monitoring service provisioning logs, network traffic variations, and user mobility patterns, AI systems can mine profound insights into macro-level user behaviors and network congestion dynamics. 
Additionally, at the physical layer, the procedures of THz channel estimation and beam tracking yield massive CSI, which intrinsically represents detailed signatures of the surrounding environment. 
Consequently, the THz communication link itself serves as a passive data collector, providing AI models with a unique, continuous stream of environmental and spatial-temporal data.

Beyond implicit data mining from communication system operations, the paradigm of ISAC empowers THz networks to capture high-fidelity physical data explicitly and intuitively. 
Benefiting from sub-millimeter wavelengths and ultra-wide bandwidths, THz signals inherently enable exceptionally high precision for sensing and environmental perception. 
Specifically, THz sensing facilitates pervasive data collection across diverse application scenarios, encompassing human activity recognition, indoor localization, gesture and gait recognition, and heart rate monitoring \cite{wang2022integrated}. By seamlessly integrating multi-modal sensing and communication, the THz infrastructure continuously harvests massive volumes of authentic, high-quality physical data. 
Consequently, THz sensing provides the indispensable data to train and optimize related AI models.

\begin{table*}[t]
\centering
\caption{Summary of THz Physical Computing for AI}
\label{tab_thz_physical_computing}
\scriptsize
\setlength{\tabcolsep}{1.7pt}
\renewcommand{\arraystretch}{1.03}
\begin{tabularx}{\textwidth}{|>{\centering\arraybackslash}m{1.3cm}|>{\centering\arraybackslash}m{4.5cm}|>{\centering\arraybackslash}m{2.8cm}|>{\centering\arraybackslash}X|>{\centering\arraybackslash}m{3.1cm}|>{\centering\arraybackslash}m{0.7cm}|}
\hline
\textbf{Paradigm} & \textbf{Physical Implementation} & \textbf{AI Function} & \textbf{Main Focus} & \textbf{Validation} & \textbf{Ref.} \\
\hline
\multirow{5}{*}[-0.1cm]{\shortstack{Free-space\\diffraction}}
& Passive diffractive layers
& Image classification
& Propagation-based inference
& Selected-sample tests
& \textbf{\cite{lin2018all}} \\
\cline{2-6}
& Diffractive front end$\&$electronic network
& Classification
& Reduced electronic input size
& Numerical study only
& \textbf{\cite{mengu2020analysis}} \\
\cline{2-6}
& Broadband diffraction$\&$spectral readout
& Classification
& Spatial-to-spectral encoding
& Selected-sample tests
& \textbf{\cite{li2021spectrally}} \\
\cline{2-6}
& Diffuser-trained layers$\&$spectral readout
& Classification
& Unseen-diffuser robustness
& Binary THz demonstration
& \textbf{\cite{bai2023all}} \\
\cline{2-6}
& Two jointly printed diffractive layers
& Binary pattern recognition
& Rotation and scaling invariance
& Transformed-pattern tests
& \textbf{\cite{xiong2024terahertz}} \\
\hline
\multirow{2}{*}[-0.15cm]{\shortstack{Guided-wave\\computing}}
& Plasmonic couplers$\&$phase-delay structures
& Image/user classification
& Planar physical inference
& Measured-response evaluation
& \textbf{\cite{gao2024terahertz}} \\
\cline{2-6}
& Dispersion-engineered plasmonic waveguides
& Parallel classification
& Frequency multiplexing of tasks
& Measured-response evaluation
& \textbf{\cite{gao2026dispersion}} \\
\hline
\end{tabularx}
\vspace{-3mm}
\end{table*}

\subsection{THz Physical Computing for AI}
The physical properties of THz waves can open up a novel paradigm for AI computing. Most existing AI computing relies on digital electronic processors to execute large-scale multiply-and-accumulate operations. As model complexity and input dimensionality continue to increase, repeated arithmetic operations, memory access, and data transfers between memory and processing units impose substantial energy and latency overheads \cite{lin2018all}. Instead, THz waves enable a physical computing paradigm in which trainable parameters are directly encoded into the phase or complex transmission responses of diffractive or guided-wave structures. By exploiting intrinsic wave phenomena such as diffraction, interference, dispersion, and EM coupling, these structures can realize desired computational transformations directly through THz wave propagation. Such computation proceeds in parallel across spatial and spectral degrees of freedom and can be implemented using passive structures without repeatedly executing digital arithmetic operations. These properties enable THz systems to move part of AI inference and feature processing from electronic processors into the EM domain, thereby potentially reducing data conversion, memory movement, and computational overhead. Table~\ref{tab_thz_physical_computing} summarizes THz physical computing studies.

A representative realization of this paradigm is the diffractive deep neural network (D$^2$NN) proposed in \cite{lin2018all}. After digital training, the learned parameters are physically encoded into cascaded passive diffractive layers, through which 0.4 THz waves perform inference by successive diffraction and interference. 
This architecture demonstrates that learned transformations can be executed directly through THz wave propagation, enabling highly parallel inference with negligible power consumption in the passive computing layers and propagation-limited processing latency.
A subsequent study in \cite{mengu2020analysis} systematically investigates the design of D$^2$NNs and further integrates a diffractive physical front end with a compact electronic neural network.
This hybrid architecture compresses high-dimensional optical inputs into low-dimensional task-relevant representations before electronic processing, substantially reducing the detector and downstream processing requirements while maintaining high classification accuracy.
Exploiting the wide spectral dimension of THz signals, the work in \cite{li2021spectrally} further maps spatial information into a low-dimensional spectral representation, enabling single-pixel machine vision and substantially reducing the amount of information that must be electronically detected and processed. 
Beyond controlled propagation conditions, the diffractive network in \cite{bai2023all} is trained across random scattering environments to maintain classification capability through previously unseen diffusers. Similarly, geometric invariance can be embedded directly into the physical transformation, as demonstrated in \cite{xiong2024terahertz}, where rotated and scaled target patterns are mapped to fixed decision regions. Therefore, these two studies extend D$^2$NN-based physical inference toward improved robustness against propagation perturbations and input transformations.

While these studies mainly rely on free-space diffraction, recent work has moved toward more compact and integrated THz computing architectures. 
The spoof plasmonic neural network (SPNN) in \cite{gao2024terahertz} replaces free-space propagation with cascaded spoof surface plasmon polariton couplers and phase-delay structures. THz surface waves therefore perform signal mixing and learned transformations within a planar circuit, providing a pathway toward integrated physical neural computing. 
This concept is further extended in \cite{gao2026dispersion} through dispersion-engineered SPNNs, where different THz frequencies experience distinct effective transformations within the same hardware. The frequency dimension can consequently support parallel AI tasks while simultaneously enabling communication functions such as frequency demultiplexing, illustrating that the wide THz spectrum can serve as both communication bandwidth and a physical computing resource.

\subsection{Lessons Learned}

\begin{itemize}
    \item \textbf{Key communication enabler.}
    THz networks can provide workload-aware communication by adapting connectivity and resources to computation dependencies in AI services. Meanwhile, THz-enabled AI systems should account for beam misalignment or other link uncertainties to maintain stable AI task performance.

    \item \textbf{Data acquisition infrastructure.}
    THz networks can generate valuable data beyond transporting it. Network operations continuously expose channel, mobility, and traffic information, while high-resolution THz sensing can directly acquire fine-grained physical-world information. Such capabilities provide rich and continuously updated data sources for environment-aware and embodied AI.

    \item \textbf{Physical computing medium.}
    THz wave propagation can directly participate in AI computation. This paradigm enables highly parallel AI processing with propagation-level latency and low energy consumption.
\end{itemize}


\section{Open Research Issues and Future Directions}
\label{sec_future}
This section discusses the open research issues and future directions for symbiotic development of AI and THz networks.

\vspace{-3mm}
\subsection{Limitations of AI for THz Networks}
Despite their advantages in complex modeling and optimization, AI methods introduce limitations that may hinder reliable deployment in practical THz networks. 
A fundamental issue is dependence on representative training data. 
High-quality THz measurements are costly to collect across diverse environments, while models trained primarily on simulated data may mismatch real propagation conditions \cite{hu2024transfer}. 
Moreover, models trained on specific datasets may experience considerable performance degradation when deployed in previously unseen environments \cite{kalor2021prediction,yu2023adaptive,xu2025scenario}. 
This dependence on scenario-specific data therefore makes generalization across heterogeneous and dynamic THz networks particularly challenging. 
Several promising approaches may help address these challenges.
Transfer learning and meta-learning reuse knowledge from related environments to reduce target-domain data and retraining requirements, while domain adaptation can further mitigate distribution mismatches between training and deployment conditions. 
Mixture-of-Experts architectures provide another direction by maintaining specialized models for different operating conditions and dynamically selecting suitable experts according to the observed environment. 
Future AI designs should further explore adaptive and robust mechanisms to maintain reliable performance across diverse environments and network conditions.

Another concern is the limited interpretability and physical consistency of data-driven models. 
Black-box neural networks learn input-output mappings without explicitly exposing the underlying communication mechanisms, which complicates performance diagnosis.
Model-driven learning, as illustrated in Sec.~\ref{sec_channel} and Sec.~\ref{sec_phy}, embeds established signal-processing or optimization procedures into trainable network architectures, allowing learned parameters to operate within predefined algorithmic and physical structures \cite{yu2023adaptive,unfolding_tsp2023}. 
This explicit correspondence between network modules and underlying domain knowledge makes the resulting models more interpretable.
Going forward, physics-informed AI \cite{bian2025genert} may further improve interpretability by embedding explicit THz-specific physical principles into model design, making learned decisions more closely grounded in the underlying system mechanisms.

As intelligence becomes increasingly embedded throughout THz networks, privacy and security have evolved into critical issues. 
AI-enabled communication, sensing, and network management continuously process information such as user locations, mobility patterns, and environmental observations, which may reveal sensitive spatial or behavioral information. 
Distributed intelligence further requires such information or learned model updates to be exchanged across terminals, edge servers, and network controllers, thereby enlarging the set of entities that must be trusted. 
Compromised data sources or unauthorized access may expose private information and corrupt observations used for network decisions. 
Future AI-enabled THz systems should integrate privacy-preserving learning with secure and trustworthy network operation.


\vspace{-3mm}
\subsection{AI-native Cross-layer Design}
As discussed in Sec. \ref{sec_network}, THz networks require cross-layer coordination for efficient end-to-end operation \cite{xia2021multi}.
However, cross-layer design introduces high-dimensional non-convex optimization problems that challenge conventional methods. Additionally, as more application scenarios emerge, including autonomous driving, generative video communications, and token transmission~\cite{liu2026tonic}, increasingly diverse QoS requirements further increase the complexity of cross-layer networking. These challenges motivate AI-native networking, which embeds intelligence into the protocol architecture and allows different layers to share network representations and coordinate decisions toward end-to-end objectives. In this holistic architecture, rigid protocol stack boundaries are dismantled and protocol overhead can be reduced, improving adaptation to changing QoS and network conditions. Therefore, AI-native cross-layer design is an important research direction for improving the efficiency and adaptability of THz networks.

\vspace{-3mm}
\subsection{THz Network for AI}
As 6G and future wireless networks evolve toward the deep integration of communication, sensing, computing, and intelligence, their role is expanding beyond data transmission. 
By coordinating multi-dimensional resources distributed across the network, such as communication, computing, and memory resources, while continuously acquiring information from the physical environment, future networks can provide an integrated infrastructure for efficient AI service delivery.
Although substantial progress has been made in AI infrastructure through the joint design of memory, computation, communication, and scheduling \cite{rajbhandari2020zero, narayanan2021megatron, kwon2023pagedattention, zhong2024distserve}, increasing AI workloads are placing growing demands on the flexibility and capacity of the underlying interconnects. 
This evolution creates an opportunity for THz networks to extend AI infrastructure and improve system-level AI capabilities.

One promising direction is to develop hierarchical THz-enabled AI infrastructures spanning accelerator interconnects, data centers, and edge networks. Within computing clusters, current THz wireless data center research mainly focuses on reconfigurable rack-to-rack links \cite{jiang2026thzsync}. 
THz on-chip communications \cite{wang2024chip} provide another promising direction for AI computing infrastructures, where reduced transmission distances can support higher data rates and more reliable connectivity. Such links can facilitate communication among GPUs and other computing units and complement high-performance interconnects such as NVLink.
However, large-scale computing clusters are costly and scarce, which limits their ability to satisfy the rapidly growing demand of large AI models. A promising alternative is to coordinate distributed edge resources to collaboratively execute AI tasks across the network. This introduces a complex orchestration problem over heterogeneous multi-dimensional resources, while the resulting AI services must still satisfy stringent QoS requirements in latency, reliability, and availability.
Moreover, distributed AI raises privacy and security concerns, as users’ requests, location information, and local data may be exposed when AI tasks are processed across multiple network and computing nodes. Future THz-based AI infrastructures should therefore jointly consider resource orchestration, service guarantees, and privacy-preserving execution as part of their system design.

\begin{figure}[t!]
    \centering
    \includegraphics[width=0.9\linewidth]{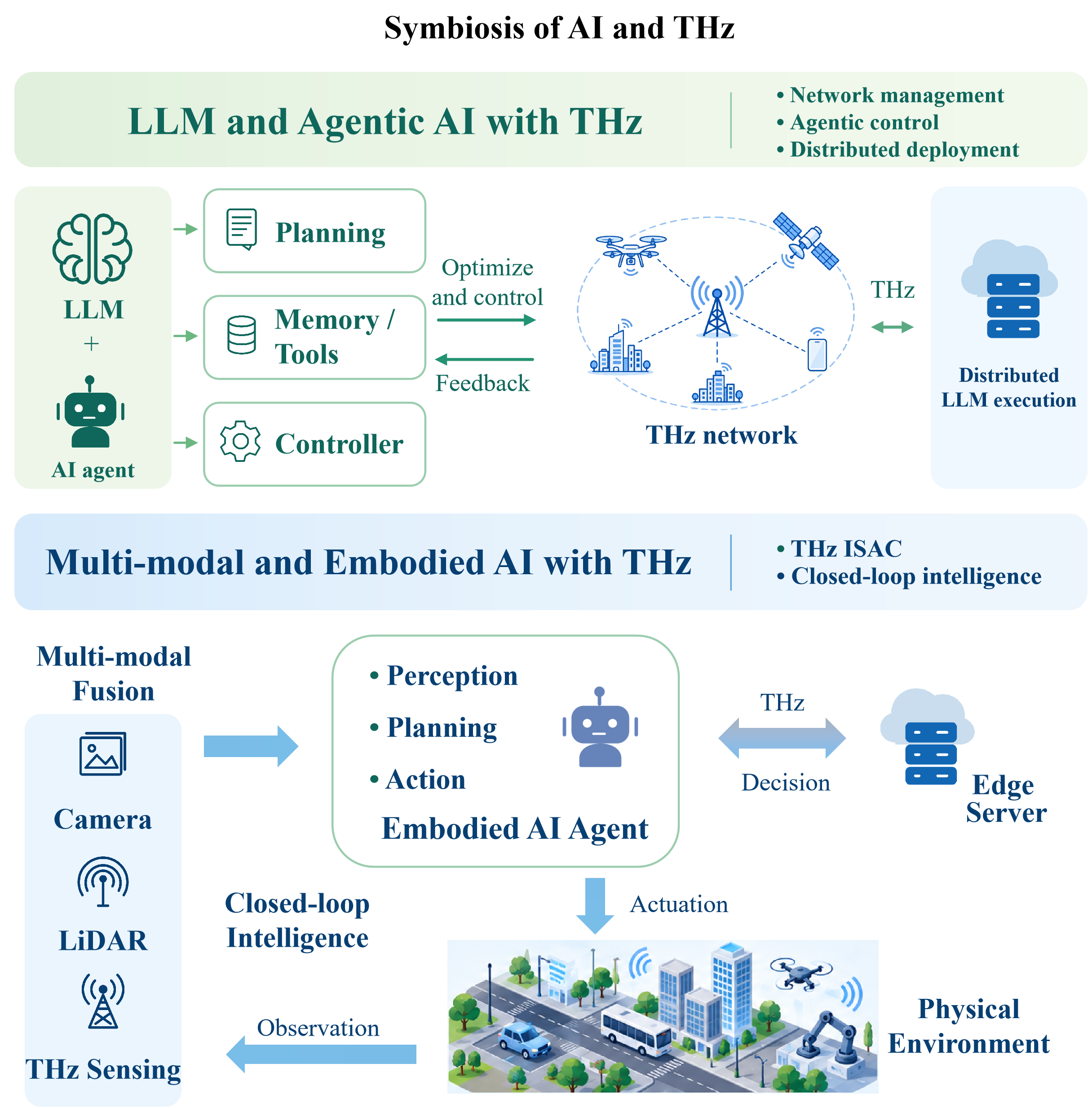}
    \caption{Future directions.}
    \label{fig_discussion}
    \vspace{-3mm}
\end{figure}

\subsection{Multi-modal Fusion}
Multi-modal fusion integrates diverse sensory modalities like vision, LiDAR, and radar to construct comprehensive environmental representations. This paradigm exhibits immense potential for ISAC \cite{lin2024environment} and channel modeling \cite{bai2025multi}. Specifically, as for ISAC, THz waves enable high-precision sensing but suffer from limited range and substantial directional scanning overhead. Multi-modal fusion mitigates these drawbacks by compensating for the coverage and efficiency drawbacks of standalone THz sensing. Similarly, relying solely on wireless signals for channel modeling fails to capture complex geometries, which causes inaccurate estimation and beam alignment.
Moreover, the THz spectrum also empowers multi-modal technologies. Unique THz scattering and reflection characteristics serve as a high-resolution modality to capture fine-grained material properties and geometric contours. Furthermore, the ultra-high bandwidth of THz communications supports the seamless exchange of enormous heterogeneous datasets generated by diverse sensory networks.
However, combining multi-modal data remains challenging due to distinct physical properties and heterogeneous structures. Directly merging raw data fails to exploit complementary strengths. Consequently, designing robust feature fusion and modality alignment mechanisms to process heterogeneous data represents a critical future direction for advancing THz networks.

\vspace{-3mm}
\subsection{LLM and Agentic AI with THz}
LLMs demonstrate remarkable potential in wireless networks due to advanced reasoning and generalization capabilities \cite{ zheng2025beamllm, long2025survey,wu2024netllm}.
A key issue for their extension to THz systems is how to represent wireless observations in a form that preserves the underlying propagation characteristics.
Existing studies provide some heuristic insights. The framework in \cite{liu2024llm4cp} introduces dedicated preprocessing and embedding modules to adapt channel sequences to a pre-trained LLM for channel prediction. 
One step further, the work in \cite{bian2026airfm} develops a wireless foundation model trained on delay-Doppler-angle domain CSI to learn transferable representations across different wireless tasks and scenarios.
These two paradigms suggest different implementation paths for THz networks. Developing THz-specific foundation models or dedicated adaptation mechanisms for general pre-trained models remains an important research direction.
Additionally, the advantage of LLMs or wireless foundation models lies in their potential to share knowledge across heterogeneous network functions. These studies \cite{liu2025llm4wm,xu2025llmnearfield,shao2024wirelessllm} have indicated that LLMs can support diverse communication tasks across both the physical and network layers.
This capability is particularly important for THz networks, where narrow beams and highly dynamic links couple tasks across the protocol stack. Cross-task and cross-layer knowledge sharing can therefore improve network management efficiency and environmental adaptability, and maintain stable end-to-end performance.

Agentic AI leverages LLMs for reasoning and task planning, and further incorporates memory, tool invocation, and action execution to enable autonomous network operation \cite{jiang2026large,tong2026wirelessagent}.
For THz networks, an agent can collect channel states, beam conditions, and blockage indicators, combine them with historical observations and domain knowledge, and invoke tools to formulate and execute feasible actions.
This tool-assisted paradigm has been explored in existing work.
In \cite{li2026agenticwireless},  an LLM-based agent translates communication intents into physical-layer decisions and executes them through external wireless tools for link configuration.
The task-driven agent in \cite{han2026dragonfly} selects vision, LiDAR, and THz sensing modules according to sensing objectives and integrates their outputs through physics-guided constraints for environment reconstruction.
Agentic AI provides a pathway toward closed-loop autonomous management of THz systems by connecting network perception, high-level reasoning, and executable control to adapt to their dynamic and complex operating conditions.

However, deploying LLMs or agents at base stations remains challenging due to their substantial computational and memory demands. A feasible approach is to adopt a hierarchical control architecture that separates large-model reasoning from time-critical network operations \cite{tang2025oranagent}. Specifically, large LLMs or agentic models can be placed at edge servers or the cloud, while base stations execute lightweight models and specialized tools for time-sensitive radio functions.
THz can then provide high-capacity connectivity for exchanging network observations, intermediate model states, and multi-modal information between distributed network nodes and remote intelligence.
Jointly designing model placement, tool execution, and THz communication for such distributed intelligence remains an important future research direction.

\vspace{-3mm}
\subsection{Physical and Embodied AI with THz}
Physical AI integrates AI models with sensing and actuation to support autonomous interaction with the physical world~\cite{gaba2026physical}. Within this broader vision, embodied AI enables intelligent agents to perceive, reason, and act within physical environments, where communication networks connect sensing, computation, and physical control  ~\cite{yang2024embodied,bariah2024ai,dong2026networkedembodied}.
This paradigm can be extended to THz networks through a closed loop in which robots or UAVs acquire multi-modal observations, including THz sensing information, and use them to construct task-relevant environmental representations ~\cite{han2026dragonfly}. Such observations can provide both the environmental context required by embodied tasks and the channel-related information needed for THz link management.

In practical implementation, complex embodied reasoning, including large-model inference, global map construction, and multi-agent planning, often exceeds the computing and memory capabilities of embodied robots and is therefore more suitable for edge or cloud execution. In contrast, robots can retain lightweight local processing to extract task-relevant features or semantic representations before transmitting them to remote computing nodes for further reasoning~\cite{li2025taskrobotics,li2026robotsexchange,chen2026physical}. THz can therefore facilitate this process by rapidly transporting high-dimensional sensing features, semantic representations, and agent states, while the resulting decisions are returned to control both the communication system and physical agents. Additionally, emerging THz hardware supporting simultaneous multi-agent communication, localization, and sensing further provides a physical basis for such networked operation~\cite{xia2026thzchip}.
This coupling is also reciprocal, as the actions of embodied agents directly reshape THz propagation conditions.
Agent mobility can first be exploited to avoid persistent blockage and recover favorable links, while more deliberate changes in position or orientation can further improve beam alignment and expose alternative propagation opportunities.
As the task and communication states evolve together, embodied control can become an integral part of THz networks and close the loop between physical action and network adaptation.

\section{Conclusion}
\label{sec_conclusion}
In this work, we present a systematic survey of the symbiotic relationship between AI and THz networks. We first establish the foundations by introducing key AI techniques and evaluation platforms for THz research, followed by AI-assisted THz hardware design and channel modeling. Building upon these foundations, we investigate \textit{AI for THz}, systematically examining how AI can address the intrinsic challenges of THz communications and networking across physical-layer functionalities, higher-layer protocols, and emerging services such as edge computing and sensing.
We then reverse the perspective to \textit{THz for AI}, exploring how the unique communication, sensing, and computing capabilities of the THz band can enable next-generation AI systems. Reconfigurable, high-capacity THz links can facilitate AI training and inference in data centers and interconnect distributed computing resources for efficient AI services. Meanwhile, high-resolution THz sensing can acquire rich physical-world information for multi-modal, embodied, and Physical AI, bridging digital intelligence with the physical environment. Beyond communication and sensing, THz waves can further serve as a physical computing medium, potentially reducing the computational overhead of AI inference. Together, these two complementary perspectives reveal a deeper AI--THz symbiosis, in which AI makes THz networks more intelligent and efficient, while THz provides the connectivity, perception, and computing capabilities required by future AI systems.


Looking ahead, we highlight open research directions toward a deeper symbiosis between AI and THz networks and draw the following key lessons. For \textit{AI for THz}, future research should move beyond the optimization of individual modules toward AI-native architectures that enable coordinated intelligence across the protocol stack. In this transition, AI-RAN provides a promising framework for cross-layer adaptation to highly dynamic THz environments and heterogeneous service requirements, while emerging paradigms such as LLMs and agentic AI may further enhance network autonomy, enabling self-optimization and self-healing THz connectivity.

Reciprocally, \textit{THz for AI} calls for exploiting the extreme connectivity, sensing, and computing capabilities of THz networks to extend AI infrastructure, orchestrate distributed communication and computing resources, and enhance the QoE of increasingly data-intensive AI services. Moreover, embodied and Physical AI provide a compelling convergence point, where THz-enabled communications and sensing allow intelligent agents to continuously sense, communicate, compute, and interact with the physical world. 
Ultimately, the evolution toward an integrated \textit{AI--THz symbiosis} may transform THz from a communication technology into an intelligent infrastructure for future AI systems.
\bibliographystyle{IEEEtran}
\bibliography{references}

\end{document}